\documentclass[aps,pra,10pt,twocolumn,amsmath,amssymb,showpacs,
    longbibliography,superscriptaddress,nofootinbib]{revtex4-1}

\usepackage{amsmath,amssymb,amsfonts} 
\usepackage{graphicx}

\usepackage{hhline}
\usepackage[latin1]{inputenc}
\usepackage{subfigure}
\usepackage{float}
\usepackage[colorlinks=true,citecolor=blue,linkcolor=blue]{hyperref}
\usepackage{bm}
\usepackage{bbm}
\usepackage{nicefrac}
\usepackage{wasysym}

\newcommand{\figheight}{10cm}
\newcommand{\eq}[1]{Eq.~(\ref{eq:#1})}
\newcommand{\eqs}[2]{Eqs.~(\ref{eq:#1}) and~(\ref{eq:#2})}
\newcommand{\equ}[1]{Equation~(\ref{eq:#1})}

\newcommand{\fig}[1]{Fig.~\ref{fig:#1}}
\newcommand{\sect}[1]{Sec.~\ref{sec:#1}}
\newcommand{\ftnt}[1]{\textsuperscript{\ref{#1}}}

\def\nn{\nonumber\\}

\def\dk{[d\kk]}
\def\im{{\rm Im}}
\def\re{{\rm Re}}

\def\beq{\begin{equation}}
\def\eeq{\end{equation}}
\def\bea{\begin{eqnarray}}
\def\eea{\end{eqnarray}}

\def\ket#1{\vert#1\rangle}
\def\bra#1{\langle#1\vert}
\def\ip#1#2{\langle#1\vert#2\rangle}
\def\me#1#2#3{\langle#1\vert#2\vert#3\rangle}
\def\wt#1{\widetilde{#1}}

\def\ww{\omega}
\def\kk{{\bm k}}

\def\qq{{\bm q}}
\def\rr{{\bm r}}

\def\BB{{\bm B}}

\def\jj{{\bm j}}
\def\E{{\mathcal E}}
\def\bE{{\bm{\mathcal E}}}
\def\ee{\hat{\bm e}}
\def\eps{\epsilon}
\def\vareps{\varepsilon}
\def\alfa{{\mathfrak a}}

\begin{document}

\title{Gyrotropic effects in 
  trigonal tellurium studied from first principles }

\author{Stepan S. Tsirkin}
\author{Pablo Aguado Puente}
 \affiliation{Centro de F{\'i}sica de
  Materiales, Universidad del Pa{\'i}s Vasco, E-20018 San Sebasti{\'a}n,
  Spain} 
\affiliation{Donostia International Physics Centre, E-20018 San
  Sebasti{\'a}n, Spain}

\author{Ivo Souza} \affiliation{Centro de F{\'i}sica de Materiales,
  Universidad del Pa{\'i}s Vasco, E-20018 San Sebasti{\'a}n, Spain}
\affiliation{Ikerbasque Foundation, 48013 Bilbao, Spain}

\date{\today}
\begin{abstract}
  We present a combined {\it ab initio} study of several gyrotropic
  effects in $p$-doped trigonal tellurium (effects that reverse
  direction with the handedness of the spiral chains in the atomic
  structure).  The key ingredients in our study are the
  $k$-space Berry curvature and intrinsic orbital
  magnetic moment imparted on the Bloch states
  by the chirality of
  the crystal structure.  We show that the observed sign reversal with
  temperature of the circular photogalvanic effect can be explained by
  the presence of Weyl points near the bottom of the conduction band
  acting as sources and sinks of Berry curvature.  The passage of a
  current along the trigonal axis induces a rather small
  parallel magnetization, which can nevertheless be detected by
    optical means (Faraday rotation of transmitted light) due to the
    high transparency of the sample.
  In agreement with experiment, we find that when infrared light
  propagates antiparallel to the current at low doping the
  current-induced optical rotation enhances the natural
  optical rotation.  According to our calculations the plane of
  polarization rotates in the opposite sense to the bonded atoms in
  the spiral chains, in agreement with a recent experiment
  that contradicts earlier reports.
\end{abstract}
\maketitle

\section{Introduction}

The spontaneous magnetization of ferromagnetic metals gives rise to
Hall and Faraday effects at $\BB=0$. These effects are termed {\it
  anomalous}, in opposition to the ordinary (linear in $\BB$) Hall and
Faraday effects in metals lacking magnetic order.  The
scattering-free or {\it intrinsic} contribution to the 
anomalous Hall conductivity (AHC) is given
by~\cite{nagaosa-rmp10,xiao-rmp10}
\begin{subequations}
\label{eq:ahc+curv}
\begin{align}
\label{eq:ahc}
\sigma^{\rm A}_{ab}&=
-\frac{e^2}{\hbar}\int\dk\sum_n
f_0(E_{\kk n},\mu,T)\eps_{abc}\Omega^c_{\kk n},\\
\label{eq:curv}
{\bm\Omega}_{\kk n}&={\bm\nabla}_\kk\times {\bm A}_{\kk n}
=-\im
\langle 
    {\bm\partial}_\kk u_{\kk n}|\times|{\bm\partial}_\kk u_{\kk n}
\rangle,
\end{align}
\end{subequations}
where ${\bm A}_{\kk n}=i\ip{u_{\kk n}}{\partial_\kk u_{\kk n}}$
  is the Berry connection, ${\bm\Omega}_{\kk n}$ is the Berry
  curvature, $E_{\kk n}$ is the band energy, $f_0$ is the
  equilibrium occupation factor,
and the integral is over the
Brillouin zone with $\dk\equiv d^3k/(2\pi)^3$.  

The possibility of inducing similar effects in nonmagnetic conductors
by purely electrical means was raised by Baranova {\it et
  al.}~\cite{baranova-oc77}, who predicted the existence of an
electrical analog of the Faraday effect in chiral conducting liquids:
a change in rotatory power caused by the passage of an electrical
current.  In the following, we shall refer to this phenomenom as
``kinetic Faraday effect'' (kFE).\footnote{Although this is a
  nonstandard designation, we find it preferable to {\it
    current-induced optical
    activity}~\cite{vorobev-jetp79,shalygin-pss12} since the effect is
  closer to Faraday rotation than to natural optical activity.  The
  name adopted here is also consistent with that of a closely-related
  phenomenom to be discussed shortly, the {\it kinetic magnetoelectric
    effect}.}  In the kFE the induced rotatory power reverses sign
with the applied electric field $\bE$, in much the same way that in
the ordinary Faraday effect it reverses sign with~$\BB$.  Althought it
has not been observed so far in liquids, the kFE was measured in a
chiral conducting crystal, $p$-doped trigonal
Te~\cite{vorobev-jetp79,shalygin-pss12}, following a theoretical
prediction~\cite{ivchenko-jetp78}.  The effect is symmetry allowed in
the 18 (out of 21) acentric crystal classes known as {\it
  gyrotropic}~\cite{belinicher-spu80}, including those for which
natural optical rotation is disallowed.

Gyrotropic crystals  also display a nonlinear optical
effect closely related to the kFE: the {\it circular
  photogalvanic effect} (CPGE). {It consists in the generation
  of} a photocurrent that reverses sign with the helicity of
light~\cite{ivchenko-jetp78,asnin-jetp78,asnin-ssc79,belinicher-spu80,sturman-book92,ivchenko-book97},
and occurs when light is absorbed {via interband or intraband
  scattering processes,}
with the latter involving virtual transitions to other
bands~\cite{ivchenko-book97}.

When impurity scattering is treated in the constant
  relaxation-time approximation, it becomes possible to identify a
contribution to the intraband CPGE associated with the Berry curvature
of the free carriers~\cite{deyo-arxiv09,moore-prl10,sodemann-prl15}.
This ``intrinsic'' contribution, proportional to the relaxation time
$\tau$, is conveniently described in terms of the following dimensionless tensor
\begin{equation}
\label{eq:D_ab}
D_{ab}=\int\dk\sum_n
\frac{\partial E_{\kk n}}{\partial{k_a}}
\Omega_{\kk n}^b
\left(-\frac{\partial f_0}{\partial E}\right)_{E=E_{\kk n}},
\end{equation}
where the index $\kk$ has been dropped for brevity.  $D$ transforms
like the gyration tensor $g$, but unlike $g$ it is always
traceless.\footnote{\label{note-traceless} After integrating \eq{D_ab}
  by parts, the trace of the tensor~$D$ can be expressed as a
  Brillouin-zone integral of the divergence of the Berry curvature,
  weighted by the occupation factor. The fact that the Berry curvature
  is divergence-free except at isolated chiral band crossings (Weyl
  points) implies that $D$ is traceless~\cite{deyo-arxiv09}.}  This
means that $D$ can only be nonzero in 16 of the 18 gyrotropic crystal
classes; the excluded classes are O and T, for which $g$ is isotropic
(its form is tabulated in Ref.~\cite{sturman-book92} for all
the gyrotropic crystal classes).

{In addition to the CPGE,} the tensor $D$ also describes a
nonlinear anomalous Hall effect (AHE)~\cite{sodemann-prl15} that can
be viewed as the low-frequency limit of the kFE. Indeed, the kFE is
governed by a tensor $\wt{D}(\ww)$ [\eq{D-tilde} below] that reduces
to $D$ at $\ww=0$.

The flow of electrical current that gives rise to the kFE generates a
net magnetization in the {gyrotropic medium}, a phenomenom known
as {\it kinetic magnetoelectric effect} (kME)~\cite{levitov-jetp85}.
{It was first proposed for} bulk chiral
conductors~\cite{ivchenko-jetp78,levitov-jetp85} {and later for}
two-dimensional (2D) inversion layers~\cite{edelstein-ssc90,aronov-jetp89}, where it has
been studied intensively~\cite{ganichev-book12}.

\begin{figure*}[t!]
\includegraphics[width=17.5cm]{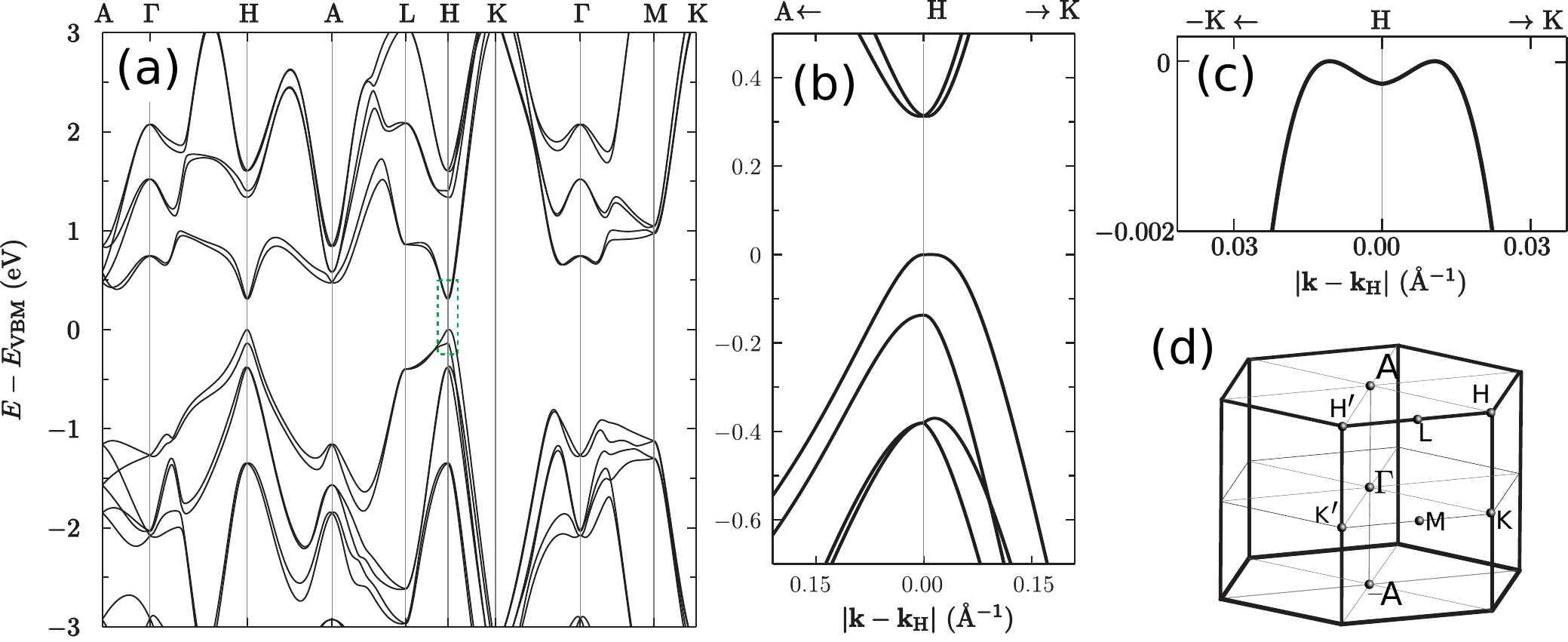}
\caption{Fully-relativistic band structure of trigonal Te, with
  energies measured from the valence-band maximum (VBM).  (b)~A
  blow-up of the region demarcated by a dashed rectangle in~(a), and
  (c)~shows the top of the upper valence band around~H, along the HK
  line.  The Brillouin zone and its high-symmetry points are displayed
  in~(d).} \label{fig:1}
\end{figure*}

A microscopic theory of the intrinsic kME effect in bulk crystals was
recently developed~\cite{yoda-sr15,zhong-prl16}.  The response,
{proportional to $\tau$,} is described by 
\beq
\label{eq:K_ab}
K_{ab}=\int\dk\sum_n\frac{\partial E_{\kk n}}{\partial{k_a}} m_{\kk n}^b 
\left(-\frac{\partial f_0}{\partial E}\right)_{E=E_{\kk n}},
\eeq
which has the same form as \eq{D_ab} but with the Berry curvature
replaced by the intrinsic magnetic moment ${\bm m}_{\kk n}$ of the
Bloch electrons.  In addition to the spin moment, ${\bm m}_{\kk n}$
has an orbital component given by~\cite{xiao-rmp10}
\beq
\label{eq:m-orb}
{\bm m}^{\rm orb}_{\kk n}=\frac{e}{2\hbar}\im
\bra{{\bm\partial}_\kk u_{\kk n}}\times
(H_\kk-E_{\kk n})\ket{{\bm\partial}_\kk u_{\kk n}},
\eeq
where we chose $e>0$.  The tensor $K$ {(with units of amperes)
  is} symmetry allowed in all 18 gyrotropic crystal classes, {and
  its}  symmetric part gives an
intraband contribution to natural optical rotation {at low
  frequencies}~\cite{zhong-prl16,ma-prb15}.

In this work, we evaluate from first principles in $p$-doped tellurium
($p$-Te) the CPGE and nonlinear AHE described by the tensor $D$, the
kFE described by $\wt{D}(\ww)$, and the kME and intraband natural
optical activity described by $K$, as well as the interband natural
optical activity.  We study them as a function of temperature and
acceptor concentration, compare with the available experimental data,
and establish correlations between them on the basis of a unified
microscopic picture. 

 The manuscript is organized as follows.  In \sect{Te:struct} we
 describe the crystal structure of trigonal Te, the energy bands, and
 the form of the gyrotropic response tensors.  In the
 subsequent sections we present and analyze our first-principles
 results for the various gyrotropic effects.  The circular
 photogalvanic effect is treated in \sect{Te:PGE}, the nonlinear
 anomalous Hall effect in \sect{Te:AHE}, the kinetic Faraday effect in
 \sect{Te:Faraday}, the kinetic magnetoelectric effect in
 \sect{Te:kME}, and natural optical activity in
   \sect{Te:NOA-intra}.  In each section, only the essential theory
 needed to understand the results under discussion is given; all
 derivations and additional technical details are left to the
 appendixes.

\section{Crystal structure, energy bands, and symmetry considerations}
\label{sec:Te:struct}

Elemental Te is a nonmagnetic semiconductor that crystalizes in two
enantiomorphic structures with space groups P$3_1$21 and P$3_2$21
(crystal class 32).  The unit cell contains three atoms disposed along
a spiral chain that is right-handed for P$3_1$21 and left-handed for
P$3_2$21, with the chains arranged on a hexagonal net.  In addition to
the screw symmetry along the trigonal axis, there are three twofold
axes lying on the perpendicular plane.

The calculations reported in this work were carried out for the
right-handed Te structure described in Ref.~\cite{asendorf-jcp57}.
For the left-handed enantiomorph, the tensors $D$ and $K$ flip sign.
These two tensors assume the forms
\beq
\label{eq:D-Te}
D=
\frac{D_\parallel}{2}
\left(
\begin{array}{ccc}
-1 & 0 & 0\\
0 & -1 & 0\\
0 & 0 & 2
\end{array}
\right)
\eeq
(note that the trace vanishes\ftnt{note-traceless})
and
\beq
\label{eq:K-Te}
K=
\left(
  \begin{array}{ccc}
    K_\perp & 0 & 0\\
    0 & K_\perp & 0\\
    0 & 0 & K_\parallel
\end{array}
\right),
\eeq
where $\parallel$ and $\perp$ denote the directions parallel and
perpendicular to the trigonal axis, respectively.  

The fully relativistic density-functional theory calculations were
done using the HSE06 hybrid functional~\cite{HSE06}.
Figure~\ref{fig:1} shows the calculated energy bands. The energy gap
of 0.312~eV at the H point is in good agreement with the value of
0.314~eV obtained with the GW method~\cite{hirayama-prl15}, and with
the experimental value of 0.323~eV~\cite{Anzin1977}.  The
characteristic ``camel-back'' shape of the upper valence band around~H
can be seen in panel~(c).  The band structure in \fig{1} is in good
agreement with other fully relativistic
calculations~\cite{peng-prb14,hirayama-prl15}.  It was calculated in
the same way as in Ref.~\cite{tsirkin-prb17}, and we refer the reader
to that work for further details.

Below room temperature, the transport and low-frequency optical
properties of weakly $p$-doped Te are governed by the upper valence
band together with the lower conduction subbands.  The conduction
subbands have an anisotropic Rashba-type spin-orbit splitting
around~H, visible in \fig{1}(b); their spin textures (not shown) are
consistent with those reported in Ref.~\cite{hirayama-prl15}.

The three band degeneracies visible in \fig{1}(b) are Weyl
points~\cite{hirayama-prl15}. Of particular interest to the present
study is the one at~H between the conduction subbands. It has positive
chirality in the right-handed structure, which means that it acts as a
source (sink) of Berry curvature in the lower (upper) subband.
Time-reversal symmetry maps the Weyl point at H onto a Weyl point of
the same chirality at H'. More generally, it sends
$(E_{\kk n},{\bm\Omega}_{\kk n},{\bm m}_{\kk n})$ to
$(E_{-\kk,n},-{\bm\Omega}_{-\kk, n},-{\bm m}_{-\kk,n})$, so that $\kk$
and $-\kk$ contribute equally to \eqs{D_ab}{K_ab}.

\section{Circular photogalvanic effect}
\label{sec:Te:PGE}

\begin{figure}[t!]
  \includegraphics[width=\columnwidth]{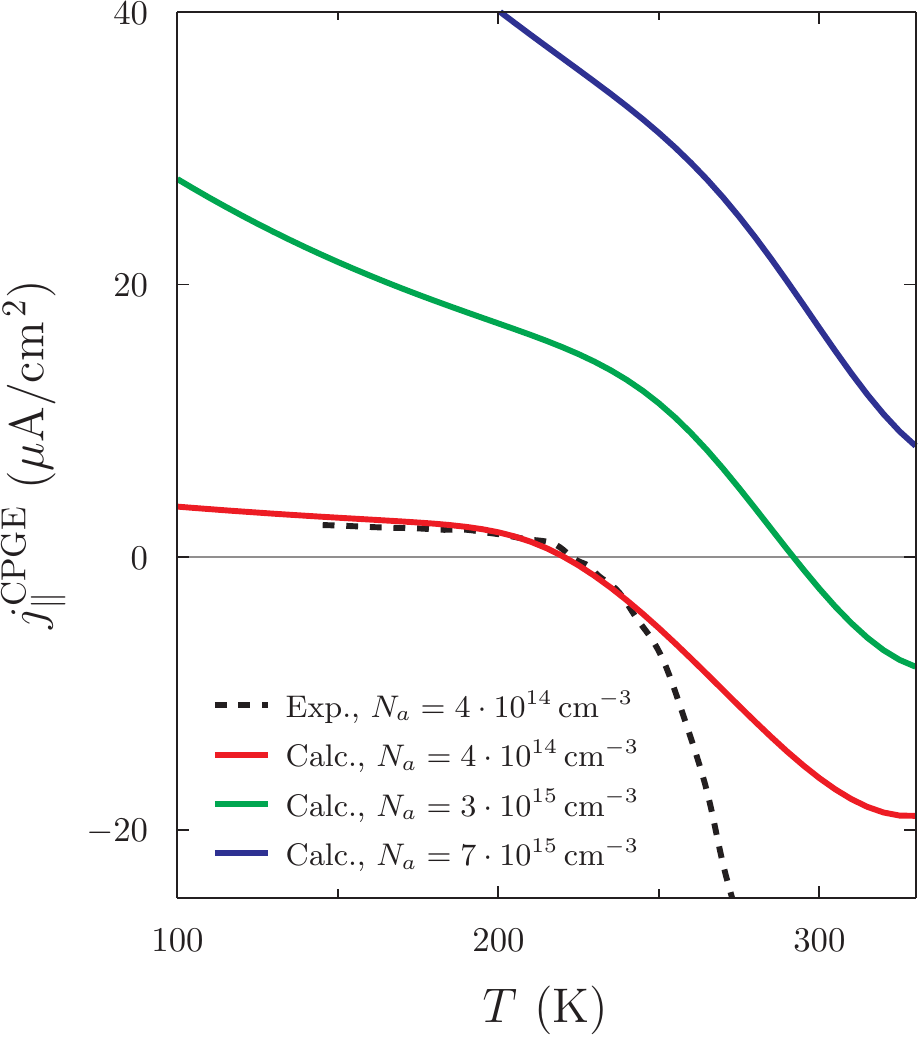}
  \caption{ (Solid lines) Temperature dependence, for different acceptor
    concentrations, of the intraband photocurrent density induced in
    right-handed Te by circularly-polarized light of positive helicity
    and intensity $I=10\,{\rm W}/{\rm cm}^2$ propagating along the
    trigonal axis in the positive direction.  According to
      \eq{cpge-Te-clean}, the photocurrent is proportional to $D_\|$.
    (Dashed line) Open-circuit photovoltage measured in
    Ref.~\cite{asnin-ssc79}, converted to a current density as
    described in the main text.
    \label{fig:2}}
\end{figure}

A detailed study of the CPGE in Te due to free-carrier 
absorption was reported in Ref.~\cite{asnin-ssc79}.  The measurements
were done at room temperature and below on samples with a residual
acceptor concentration $N_a\approx 4\cdot10^{14}$~cm$^{-3}$, using a
CO$_2$ laser source with frequency $\hbar\ww=0.117$~eV.  Under these
conditions the relaxation time exceeds $10^{-12}$~s~\cite{peng-prb14}
so that $\ww\tau\gg 1$, and \eq{cpge-Te} for the 
intrinsic contribution to the {intraband} photocurrent
becomes
\beq
\label{eq:cpge-Te-clean}
j^{\rm CPGE}_{\parallel}(N_a,T)= {\rm sgn}(q_\parallel) \left(
  2\pi\alfa {\cal P}_{\rm circ} D_\parallel \right)
\frac{e I_0}{\hbar\ww}.  \eeq The quantity
$D_\parallel(\mu(N_a,T),T)$ is given by \eqs{D_ab}{D-Te},
$\alfa\approx 1/137$ is the fine-structure constant, and~$I_0$
is the intensity of incident light with wavevector $q_\|$ and
degree of circular polarization ${\cal P}_{\rm circ}$ propagating
along the trigonal axis.  

\begin{figure*}[t!]
\includegraphics[height=\figheight]{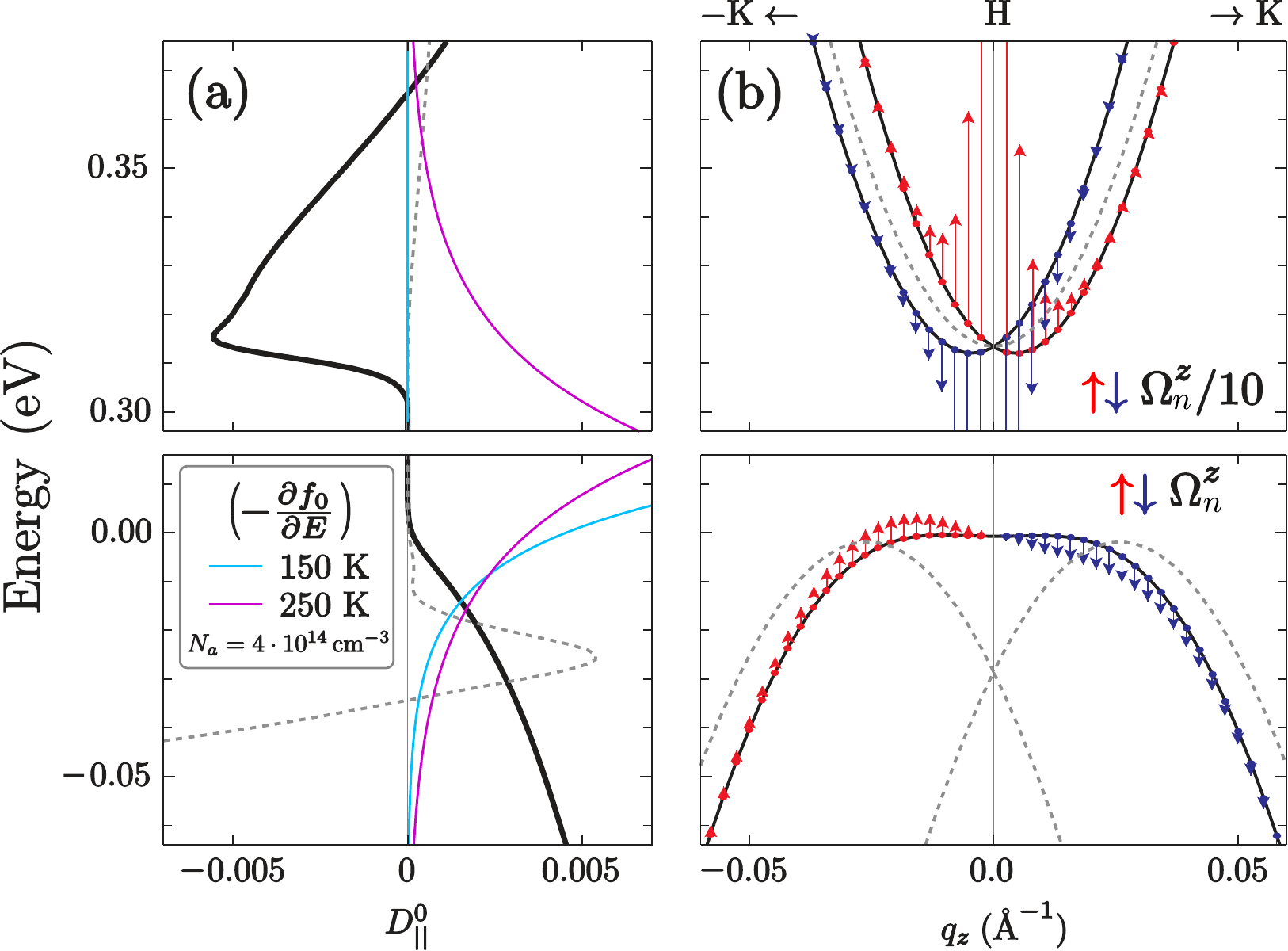}
\caption{Microscopic mechanism of the intraband circular photogalvanic 
    effect in right-handed Te.
(a) The quantity $D_\|^0$ in \eq{D-0} versus $\varepsilon$ measured
  from the VBM, calculated using a Fermi
  smearing of $23$~K with (heavy black solid line) and without (dashed
  gray line) spin-orbit coupling. The light colored lines show the
  function $-\partial f_0(E,\mu(N_a,T),T)/\partial E\vert_{E=\varepsilon}$
  plotted versus $\varepsilon$ at fixed $N_a$ and two different temperatures,
  as detailed in the inset.  (b) (Solid lines) Fully relativistic band
  structure in the vicinity of the H point. $q_z$ denotes $k_z$
  measured from the~H point along HK, and the arrows denote the
  $z$ component of the Berry curvature on each band. (Dashed lines)
  Scalar-relativistic band structure; for comparison purposes, the
  band edges have been aligned with those of the fully relativistic
  calculation.
  \label{fig:3}}
\end{figure*}

The photocurrent density calculated from \eq{cpge-Te-clean} with
$\text{sgn}(q_\|)>0$ and ${\cal P}_{\rm circ}=+1$ is plotted versus
temperature in \fig{2} for several acceptor concentrations, assuming a
laser intensity of $10\,{\rm W}/{\rm cm}^2$ (see below). The
photocurrent starts out positive at low temperature, and becomes
negative at around room temperature (except at the highest doping
level).  Such a sign reversal was indeed observed
experimentally~\cite{asnin-ssc79}. For a more detailed
  comparison, we have converted the open-circuit photovoltage and
longitudinal conductivity measured in Ref.~\cite{asnin-ssc79} into a
current density, shown as a dashed curve after an overall sign change
(the handedness of the sample was not determined in
Ref.~\cite{asnin-ssc79}).  Since the laser intensity was also not
reported, we fixed the value of~{$I_0$} in \eq{cpge-Te-clean} by
matching the experimental values at low temperature.  At the
experimental doping level the calculated photocurrent changes sign at
around 220~K, {in good agreement with
  experiment}.
 
In order to understand the temperature dependence, it is convenient to
express the quantity $D_\|$ in \eq{cpge-Te-clean} as
\begin{subequations}
\label{eq:D-conv}
\begin{align}
\label{eq:D-T}
D_\|(\mu,T)&=\int_{-\infty}^{+\infty}d\varepsilon
\, D^0_\|(\varepsilon)
\left(-\frac{\partial f_0(E,\mu,T)}{\partial E} \right)_{E=\varepsilon},\\
\label{eq:D-0}
D^0_\|(\varepsilon)&=\frac{1}{(2\pi)^3}\sum_n\int_{E_{\kk n}=\varepsilon}dS\,
\hat{v}_{\kk n}^z\Omega_{\kk n}^z,
\end{align}
\end{subequations}
where $D^0_\|(\varepsilon)\equiv D_\|(\varepsilon,T\approx 0)$, and
$\hat{\bm v}_{\kk n}$ is the unit vector along the band
velocity.\footnote{\equ{D-conv} is also convenient for numerical work.
  Once $D^0_\|(\varepsilon)$ has been calculated from \eq{D-0},
  \eq{D-T} can be used to evaluate $D_\parallel$ as a function of $T$
  and $N_a$ at a low computational cost.  The temperature dependence
  of the chemical potential is calculated assuming that at the
  temperatures of interest all dopant levels are activated.
The same approach will be
  used in subsequent sections to evaluate the tensors
  $\tilde{D}(\omega)$, $K$, and $C$ [\eq{C_ab}].}
    
Figure~\ref{fig:3}(a) shows that $D_\|^0$ has opposite signs at the
two band edges, increasing slowly into
the valence band and rapidly into the conduction band, where it peaks.
At the experimental doping level, $-\partial f_0/\partial E$ at
$150$~K is non-negligible in the valence band only, resulting in a
positive $D_\|$. At $250$~K the chemical potential $\mu$ approaches
the center of the gap, and $-\partial f_0/\partial E$ reaches the
conduction band. $D_\|$ now collects contributions of opposite signs
from the two band edges; the largest one comes from the $D^0_\|$ peak
in the conduction band, which renders $D_\|$ negative.  (When $N_a$ is
increased to $7\cdot10^{15}$~cm$^{-3}$, $\mu$ stays close to the
valence-band edge even at room temperature. The photocurrent is then
dominated by hole-like carriers, and it remains positive over the
entire temperature range of \fig{2}.)

The behavior of $D_\|^0(\varepsilon)$ at the two band edges can
be understood by inspecting the energy bands and their Berry
curvatures along the HK line [\fig{3}(b)]. Because of twofold symmetry
about $\rm\Gamma K$, $v_{\kk n}^z$ and $\Omega_{\kk n}^z$ are both odd in $q_z=k_z-{k_H,z}$,
so that $q_z$ and $-q_z$ contribute equally to \eq{D-0}.  Regarding
the $\Omega_{\kk n}^z$ profiles, note that the Berry curvature of a band
arises from its coupling to other bands [see \eq{curv-w}], and that
this coupling becomes resonantly enhanced at (near)
degeneracies~\cite{nagaosa-rmp10,xiao-rmp10}.  At the nondegenerate
valence-band edge this coupling has no singularities and as a result
$\Omega_{\kk n}^z$ varies smoothly with $q_z$, vanishing at $q_z=0$.  Apart
from a small region between the ``camel humps'' that gives a
negligible contribution, $\hat{v}_{\kk n}^z$ and $\Omega_{\kk n}^z$ have the same
sign, which explains the steady increase in $D^0_\|$ towards positive
values as $\varepsilon$ enters the valence band.

At the edge of the conduction band the Berry curvature is dominated by
the strong intersubband coupling near the Weyl point, which acts as a
monopole of Berry curvature leading to $\Omega_{\kk n}^z\propto\pm q_z^{-2}$
for small $|q_z|$~\cite{xiao-rmp10}.  When $\varepsilon$ is slightly
above the crossing energy, the two subbands give competing
contributions to \eq{D-0}: $\vert\Omega_{\kk n}^z\vert$ is larger on the
inner branch, but the outer branch has a larger energy isosurface.
For an isotropic three-dimensional (3D) Rashba model these two contributions would cancel
out,\ftnt{note-traceless} but the anisotropy of the Rashba splitting
in Te is such that the inner branch dominates the integral in
\eq{D-0}, producing a negative peak in $D_\|^0$ near the Weyl-point
energy.

\begin{figure}[t!]
\includegraphics[width=\columnwidth]{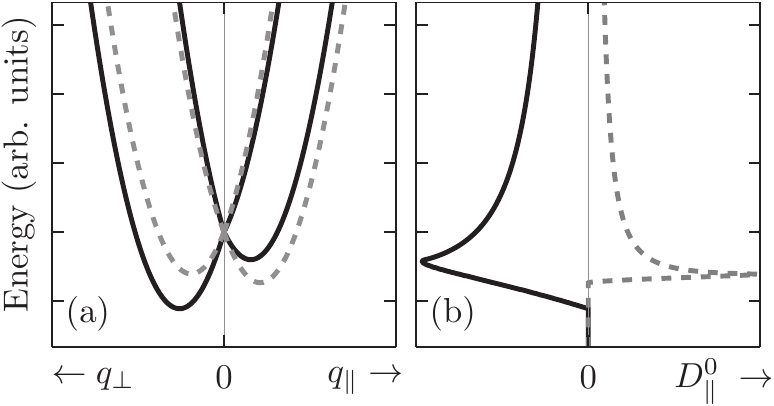}
\caption{(a) Energy bands for the anisotropic 3D Rashba model of
  \eq{Rashba} with $m_\|/m_\perp=1$, and $v_\|/v_\perp=0.6$ (solid
  line) or $v_\|/v_\perp=1.1$ (dashed line). The isolated degeneracy
  is a chiral Weyl point.  (b)~The quantity $D^0_\|$ in \eq{D-0}
  evaluated for the same choices of parameters with ${\rm sgn}(v_\|)>0$
  (positive chirality).  All axes are in arbitrary
  units. \label{fig:4}}
\end{figure}

A minimal model for the conduction-band edge is
~\cite{ivchenko-spss75}
\beq 
H^{\rm
    R}(\mathbf{q})
=\frac{\hbar^2q^2_\|}{2m_\|}+\frac{\hbar q^2_\perp}{2m_\perp}
+\hbar v_\|q_\|\sigma_z+\hbar v_\perp(q_x\sigma_x+q_y\sigma_y),
\label{eq:Rashba}
\eeq
where ${\bm q}=\kk-\kk_{\rm H}$. We have evaluated \eq{D-0}
numerically for this two-band model, starting from the analytic
expression for the Berry curvature~\cite{asboth-book16}. As expected
$D^0_\|$ vanishes in the isotropic limit, and when either
$m_\|\ne m_\perp$ or $v_\|\ne v_\perp$ a peak develops around the Weyl
crossing.  For a given chirality, the peak can change sign depending
on the ratios $m_\|/m_\perp$ and $v_\|/v_\perp$, as illustrated in
\fig{4}.

While spin-orbit coupling is not needed to generate Weyl points and
Berry curvatures in the bands of Te (in contrast to
  centrosymmetric collinear ferromagnets, where it is essential), the
intraband CPGE would be very different in its absence.  The
spin-orbit-free $D_\|^0$ and energy bands are shown as dashed gray
lines in \fig{3}. The $D_\|^0$ peak in the conduction band has been
suppressed, and a new peak has appeared in the valence band,
again associated with a Weyl crossing at~H.

In conclusion, the {intrinsic} CPGE of $p$-Te is
strongly affected by the presence of spin-orbit-induced Weyl points
at~H and~H' near the bottom of the conduction band. The large Berry
curvature around those chiral band crossings causes a sign reversal of
the photocurrent {upon cooling a weakly $p$-doped sample,}
 in agreement with experiment~\cite{asnin-ssc79}.

We emphasize that the Berry-curvature mechanism for the intraband CPGE
is different from the one discussed in Ref.~\cite{asnin-ssc79}. It
involves elastic scattering from impurities rather than inelastic
phonon scattering, and it relies on the spin-orbit splitting of the
conduction subbands that was neglected in that work.

\section{Nonlinear anomalous Hall effect}
\label{sec:Te:AHE}

\begin{figure}[t!]
\includegraphics[width=\columnwidth]{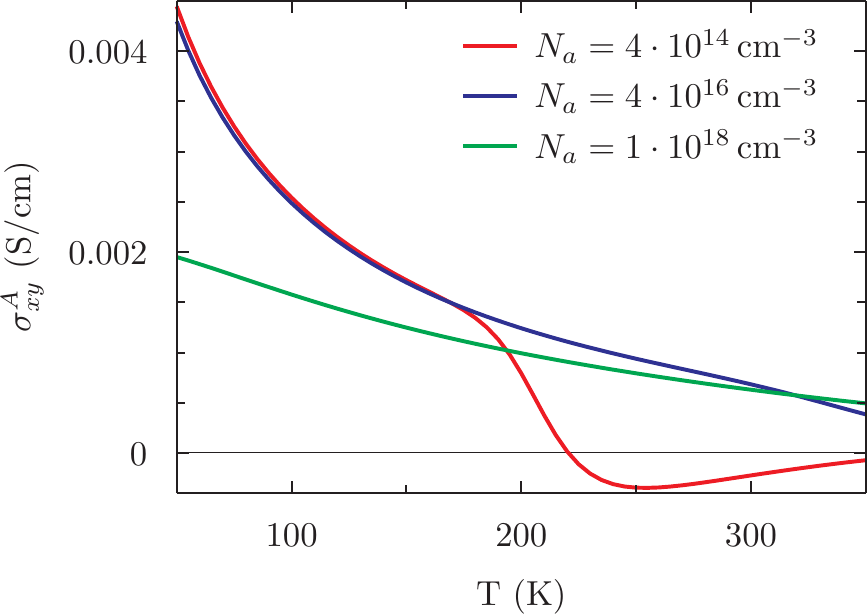}
\caption{Anomalous Hall conductivity induced in right-handed Te by a current
density  $j_\|=1000\text{ A/cm}^2$ [\eq{AHC-j}],
  plotted versus temperature at 
  different acceptor concentrations.
  \label{fig:5}}
\end{figure}

In tellurium, the nonlinear AHE takes the form of an in-plane linear
AHE proportional to the current density flowing along the trigonal
axis.  Taking $j_\|=1000\,\text{A}/\text{cm}^2$ as a reference
value~\cite{vorobev-jetp79,shalygin-pss12}, the current-induced AHC is
given by [see \eq{sigma-AH-eff}]
\beq
\label{eq:AHC-j}
\sigma^{\rm A}_{xy}(j_\parallel=1000\,\text{A}/\text{cm}^2)\approx
\frac{0.116D_\parallel}{C_\parallel\,(\text{A}/\text{cm})}
\,\text{S/cm}.
\eeq 

\begin{figure}[t!]
\includegraphics[width=\columnwidth]{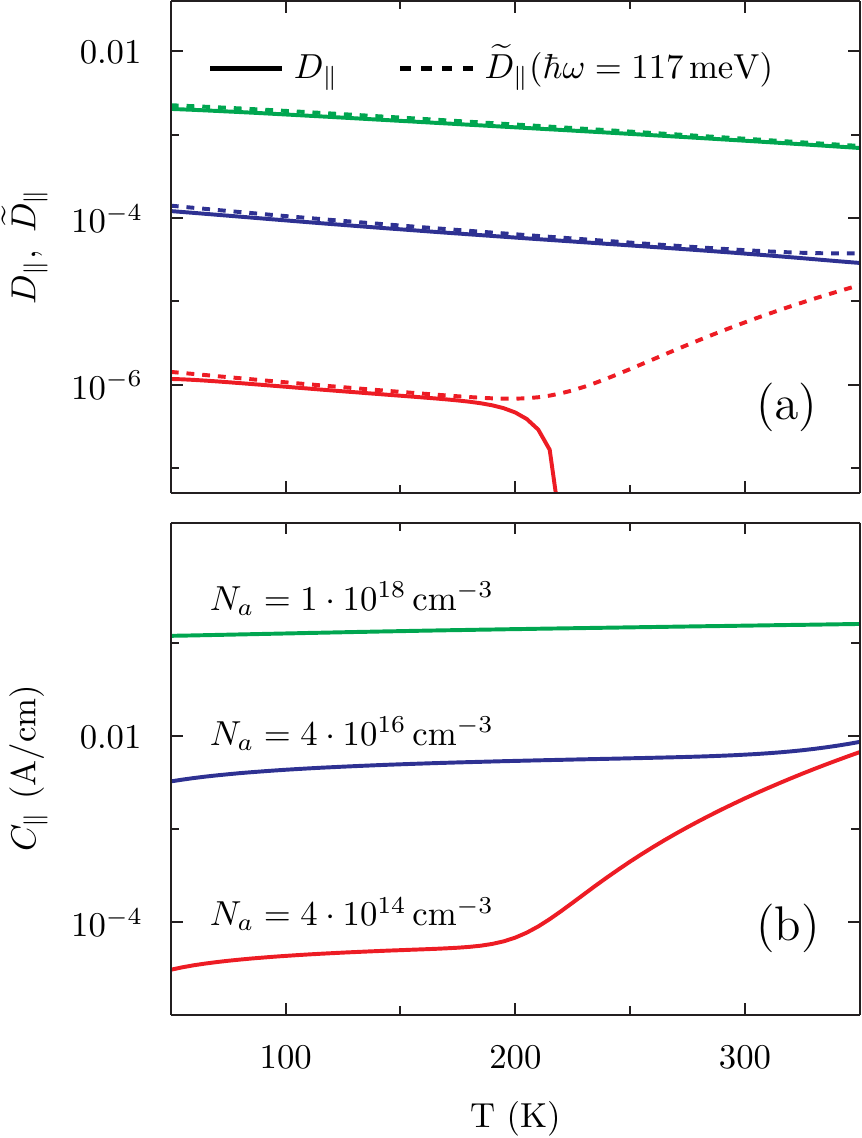}
\caption{(a) The quantities $D_\|$ [\eq{D_ab}] and $\wt D_\|$
  [\eq{D-tilde}] in right-handed Te, plotted versus
  temperature on a semilogarithmic scale for different acceptor
  concentrations. The strong dip in $\log\,D_\|$ around {220~K} at
  $N_a=4\cdot 10^{14}$~cm$^{-3}$  signals the sign change in $D_\|$ seen in
  \fig{2}.  (b) The quantity $C_\|$ [\eq{C_ab}].
  \label{fig:6}}
\end{figure}

The AHC calculated from \eq{AHC-j} is plotted versus temperature in
\fig{5} at three different doping levels.  At high doping it decreases
monotonically with temperature, while at low doping it drops to
negative values above {220~K} (due to the sign change
in $D_\|$ discussed in the previous section) and then approaches zero
from below.  Between 50
 and 170~K, the AHC is only weakly dependent on
$N_a$ over a wide doping range. This is due to a near cancellation
between the strong dependencies of $D_\|$ and $C_\|$ on $N_a$ (see
\fig{6}).

The current-induced AHC displayed in \fig{5} does not exceed
$5\cdot 10^{-3}\,\text{S/cm}$ , which is {probably} too small to
be detected (it is five orders of magnitude smaller than the
spontaneous AHC of bcc Fe~\cite{nagaosa-rmp10}). Nevertheless, the
associated Faraday rotation has been observed in the
infrared~\cite{vorobev-jetp79,shalygin-pss12}.  The analysis of that
effect will occupy us in the next section.

\section{Kinetic Faraday effect}
\label{sec:Te:Faraday}

So far, $p$-Te is the only material for which the kFE has been
measured.  The first observation was reported in
Ref.~\cite{vorobev-jetp79}, and new measurements were taken in
Ref.~\cite{shalygin-pss12}.  These works established that the
current-induced change in rotatory power ($\Delta\rho$) is linear in
$j_\|$ up to at least $\pm 1500\text{ A/cm}^2$, and that $\Delta\rho$
has the opposite (same) sign as the natural rotatory power $\rho_0$
when light travels parallel (antiparallel) to the current.

We have calculated $\Delta\rho$ from the following expression, derived
in Appendix~\ref{sec:eps-j},
\beq\label{eq:rho-j-main}
\Delta\rho(\ww,j_\parallel)={\rm sgn}(q_\parallel)
\frac{\alfa\wt{D}_\parallel(\ww)j_\parallel}{n_\perp(\ww)C_\parallel}
\eeq
(our sign convention for optical rotation is specified in
Appendix~\ref{sec:optical-rotation-phen}).  Here $q_\|$ is the
wavevector of light, and $n_\perp$ is the index of refraction; we used
{the value} $n_\perp=5.15$ calculated from \eqs{n-perp}{sigma-S},
which is slightly higher than the experimental value of
4.8~\cite{refractive-Te}. When light travels parallel to the current
[$\text{sgn}(q_\|)=\text{sgn}(j_\|)$], $\Delta\rho$ has the same sign
as the quantity $\wt{D}_\parallel(\ww)$ defined by
\beq
\label{eq:D-tilde}
\wt{D}_{ab}(\ww)=\int\dk\sum_n
\frac{\partial E_{\kk n}}{\partial{k_a}}\wt\Omega^b_{\kk n}(\ww)
\left(-\frac{\partial f_0}{\partial E}\right)_{E=E_{\kk n}},
\eeq
a finite-frequency generalization of \eq{D_ab} obtained by
  replacing ${\bm\Omega}_{\kk n}$ therein with
  $\wt{\bm\Omega}_{\kk n}(\ww)$ given by \eq{curv-w}.

  In addition to $\Delta\rho$, we have calculated the rotatory power
  $\rho_0$ caused by the natural optical activity of Te at $j_\|=0$.
  We used the formalism described in Appendix~\ref{sec:NOA-inter} to
  evaluate $\rho_0$ ignoring the influence of doping (the effect of
  doping on $\rho_0$ will be analyzed in \sect{Te:NOA-intra}, where it
  is shown to be negligible at the doping levels used in the kFE
  measurements~\cite{vorobev-jetp79,shalygin-pss12}).

\begin{table}
  \caption{\label{table:rotatory} Natural rotatory power (in units of rad/cm), and {current-induced}
    change in rotatory power divided by the current density
    (in units of $10^{-5}\text{rad}\cdot\text{cm}/\text{A}$) at $\hbar\ww=0.117$~eV and $T=77$~K for two different
    doping concentrations.
    The sign of
    $\Delta\rho/j_\|$ corresponds to  light propagating in the positive
    direction along the trigonal axis [${\rm sgn}(q_\|)>0$ in \eq{rho-j-main}].
  }
\begin{tabular}{cccc}
  \hline\hline
  & $\rho_0$ & $\Delta\rho/j_\|$  & Handedness\\
  \hline
  Expt. & $1.57\pm 0.03$\footnote{Ref.~\cite{fukuda-pss75}, undoped samples.} & 
$-9.5\pm 0.4,\footnote{Ref.~\cite{shalygin-pss12}, $p$-doped samples with $N_a=4\cdot 10^{16}\text{ cm}^{-3}$.}\,
-6\footnote{Ref.~\cite{vorobev-jetp79}, $p$-doped samples with $N_a=1.5\cdot 10^{17}\text{ cm}^{-3}$.}$& Unknown\\
  Theory & -0.86 & 4.5,\, 4&  Right-handed\\
  \hline\hline
\end{tabular}
\end{table}

Table~\ref{table:rotatory} shows the calculated values of $\rho_0$ and
$\Delta\rho/j_\|$ alongside the experimental ones, measured on samples
of unknown handedness. In agreement with experiment, we find that
$\Delta\rho$ has the opposite sign from $\rho_0$ when light travels
parallel to the current (we defer the discussion of absolute signs to
Sec.~\ref{sec:signs}).
The calculated $|\rho_0|$ and $|\Delta\rho|$ are smaller by roughly a
factor of two compared to the measured values, which can be considered
a fair level of agreement.  The calculated $|\Delta\rho|$ decreases
only slightly as $N_a$ is increased from $4\cdot 10^{16}$ to
$1.5\cdot 10^{17}\text{ cm}^{-3}$.  The larger decrease seen in the
experimental values was attributed in Ref.~\cite{shalygin-pss12} to
technical differences relative to Ref.~\cite{vorobev-jetp79}.

At $j_\|=1000\text{ A/cm}^3$, $\Delta\rho$ is about five orders of
magnitude smaller than the spontaneous Faraday rotatory power of bcc
Fe~\cite{coren:jpa-00205745}.  {This is the same difference in
  orders of magnitude that was found in the previous section for the
  AHC. However, the smallness of the kFE} is compensated by the high
transparency of Te in the infrared, which allows one to measure the
optical rotation across a cm-sized
sample~\cite{vorobev-jetp79,shalygin-pss12}, compared to
$\sim10^{-6}$~cm-thick iron films~\cite{coren:jpa-00205745}.

\subsection{Doping and temperature dependence}
\label{sec:Faraday-T}

\begin{figure}[t!]
\includegraphics[width=\columnwidth]{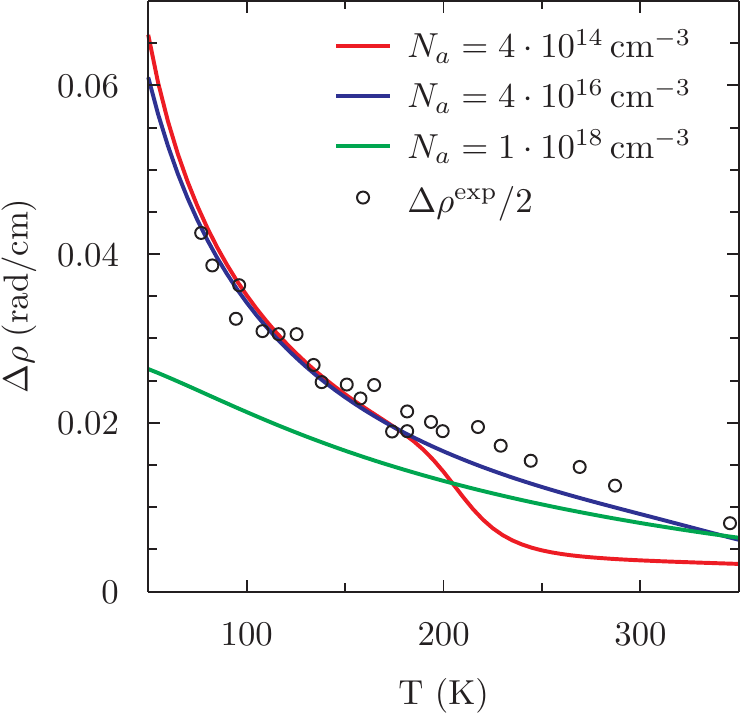}
\caption{Temperature dependence of the change in the rotatory power of
  right-handed Te induced by a current density of
  $1000\text{ A/cm}^2$. The optical frequency is $\hbar\ww=0.117$~eV,
  $N_a$ is the doping level, and the sign of $\Delta\rho$ corresponds
  to light propagating parallel to the current along the trigonal
  axis. {The open circles denote} experimental data
  \cite{shturbin-semicond1981,shalygin-thesis} {taken at}
  $N_a=3.2\cdot10^{16}\,\text{cm}^{-3}$, {which has been rescaled by a
    factor of $1/2$ for comparison purposes.}
    \label{fig:7}}
\end{figure}

Figure~\ref{fig:7} shows a weak doping dependence of $\Delta\rho$ at
low doping between 50 and 170~K, in good agreement with the
experimental data in Ref.~\cite{shalygin-thesis} (p.~27), and a
monotonic decrease with temperature.  The decrease is by a factor of
three to four between 77 and 300~K, in agreement with an earlier
theoretical estimate~\cite{averkiev-1984}. Apart from {the
  previously mentioned} overall factor of two which at present we
cannot account for, the calculated $\Delta\rho$ agrees rather well
with the experimental data reported in
{Refs.}~\cite{shturbin-semicond1981} and~\cite{shalygin-thesis}
(p.~35), as indicated by the open circles in \fig{7}.

Even at the lowest doping, $\Delta\rho$ {shows} no sign change
(only a dip) around  {220~K}. {This behavior,
  which is in} contrast to the CPGE and the nonlinear AHE, can be
understood from \fig{6}(a) where at
$N_a=4\cdot 10^{14}\text{ cm}^{-3}$ the quantity $\wt D_\|$ maintains
its sign as $T$ goes above  {220~K}, whereas $D_\|$
changes sign.

\begin{figure*}[t!]
\includegraphics[width=12.5cm]{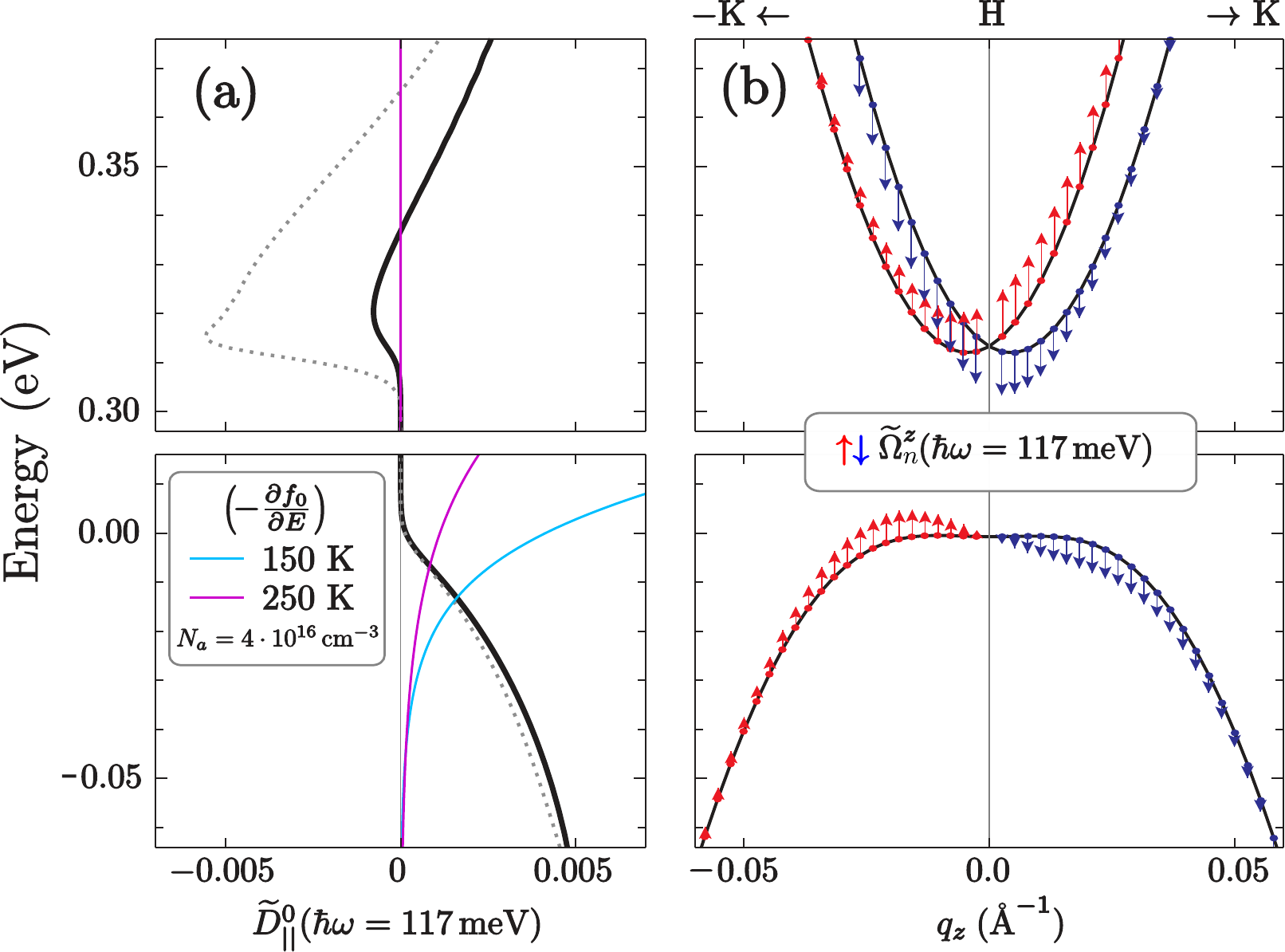}
\caption{Microscopic mechanism of the kinetic Faraday effect in
  right-handed Te at $\hbar\ww=0.117\text{ eV}$. The figure is similar
  to \fig{3}, but with $D_\|^0$ replaced by $\wt D^0_{ab}(\ww)$ [the
  low-temperature limit of \eq{D-tilde}], $\Omega^z_{\kk n}$ by
  $\wt\Omega^z_{\kk n}(\ww)$ [\eq{curv-w}], and a different doping
  level when plotting $-\partial f_0/\partial E$ in (a).  The dotted
  line in (a) represents $D_\|^0$, and is identical to the heavy solid
  line in \fig{3}(a).
  \label{fig:8}}
\end{figure*}

How close $\wt D_\|(\ww)$ is to $D_\|$ at a given temperature and
doping level depends on how close $\wt\Omega_{\kk n}(\ww)$ is to $\Omega_{\kk n}$
in the relevant energy bands, which in turn depends on how $\ww$
compares with $\ww_{mn}$ for the dominant transitions in
\eq{curv-w}. We proceed as in Sec.~\ref{sec:Te:PGE}, expressing
$\wt D_\|(\ww)$ in terms of $\wt D^0_\|({\vareps,}\ww)$ according
to \eq{D-conv}.  {The band-edge behavior of
  $\wt D^0_\|(\vareps,\ww)$ and $\wt\Omega^z_{\kk n}(\ww)$ at
  $\hbar\ww=0.117\text{ eV}$} is depicted in \fig{8}, to be compared
with \fig{3}.  In the valence band $\wt\Omega^z_{\kk n}\approx \Omega^z_{\kk n}$
and $\wt D^0_\|\approx D^0_\|$, because the dominant coupling is with
the conduction bands that are separated by more than 0.3~eV (the
coupling to the valence band below, which is closer in energy, is
suppressed by selection rules~\cite{shalygin-pss12}). In contrast
$\vert\wt\Omega^z_{\kk n}\vert\ll\vert\Omega^z_{\kk n}\vert$ in the conduction
bands, because $0.117$~eV is a large energy compared to the Rashba
splitting of the coupled subbands. The peak in $\wt D^0_\|$ is
therefore strongly reduced compared to the peak in $D^0_\|$, and this
is the reason for $\wt{D}_\|$ not changing sign with temperature at
low doping in \fig{6}(a).

In conclusion, at the CO$_2$ laser frequency the kFE is dominated by
contributions that to a good approximation can be expressed in terms
of the Berry curvature at the top of the valence band. Since this is
the same quantity that governs the {intrinsic} CPGE at low
temperatures (\sect{Te:PGE}), one can correlate the sign of
$\Delta\rho$ with that of the photocurrent measured on the same
sample.  When linearly-polarized light travels parallel to the current
[$\text{sgn}(q_\|)j_\|>0$ in \eq{rho-j-main}], $\Delta\rho$ has the
same sign as the photocurrent induced at low temperatures by light of
positive helicity traveling in the positive direction
[$\text{sgn}(q_\|){\cal P}_{\rm circ}>0$ in \eq{cpge-Te-clean}].

\subsection{Frequency dependence}

\begin{figure}[t!]
\includegraphics[width=\columnwidth]{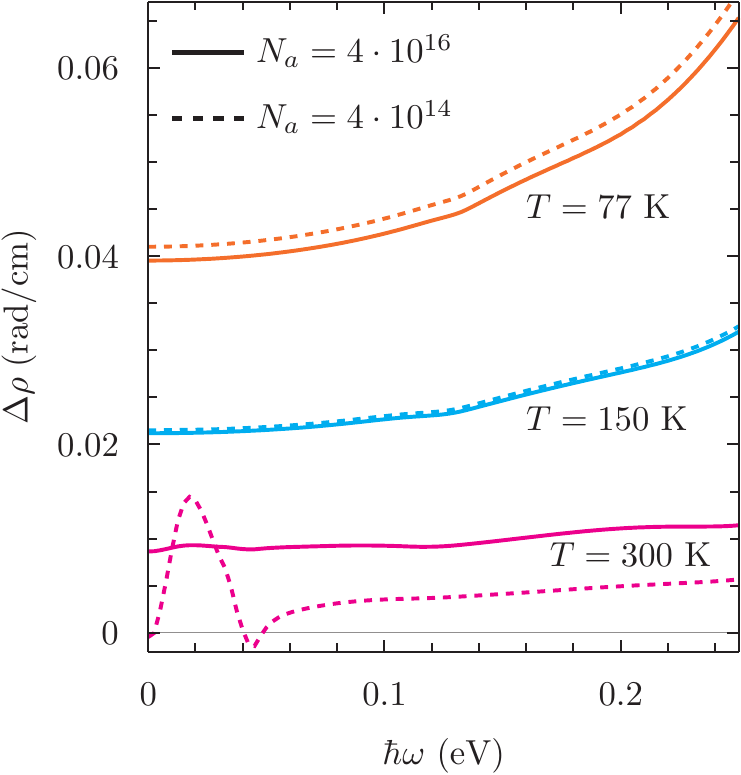}
\caption{Frequency dependence of the change in rotatory power induced
  in right-handed Te by a current density $j_\|=1000\text{ A/cm}^2$,
  at different temperatures and doping levels. In order to avoid
  singularities in \eq{curv-w} at $\ww^2=\ww_{\kk mn}^2$, $\Delta\rho$
  is calculated at complex frequencies using $\mathrm{Im}[\hbar\ww]=1$
  meV.
    \label{fig:9}}
\end{figure}

The spectral dependence of the kFE was investigated in
Ref.~\cite{shalygin-pss12} by taking additional measurements with a CO
laser, which generates radiation of higher frequency than the CO$_2$
laser. These measurements were again taken at 77~K on samples with
$N_a\approx 4\cdot 10^{16}\text {cm}^{-3}$.

Between $\hbar\ww=0.117$~eV and $\hbar\ww=0.23$~eV, $\Delta\rho$ was
found to increase by a factor of $1.7$. This is significantly less
than the increase by a factor of 4.7 in $\rho_0$~\cite{fukuda-pss75},
confirming that current-induced optical rotation and natural optical
activity are separate physical effects~\cite{shalygin-pss12}.  Our
calculated $\Delta\rho$ and $\rho_0$ increased by factors of 1.4 and
5.6 respectively over the same spectral range, in reasonable agreement
with the observed trends.

The calculated $\Delta\rho(\ww)$ is plotted in \fig{9} at different
doping levels and temperatures. At $N_a=4\cdot 10^{16}\text{ cm}^{-3}$
the spectral dependence is smooth, becoming weaker as the temperature
increases. The reason is that at this relatively high doping
$\wt D_\|$ is mostly determined by $\wt\Omega_{\kk n}$ at the valence-band
edge, which depends only weakly on frequency over the subgap spectral
range of \fig{9}.

Reducing $N_a$ to $4\cdot 10^{14}\text{ cm}^{-3}$ has practically no
effect on the spectral dependence of $\Delta\rho$ in \fig{9} at
low temperatures, since $\wt D_\|$ still originates mostly from the
top of the valence band.  At 300~K, the contribution from
the conduction bands has become significant at this low
doping. At frequencies higher than 0.05~eV this leads to a
reduction in $\Delta\rho$, due to the opposite signs of $\wt D_\|^0$
at the two band edges [\fig{8}(a)]. Below that frequency, the
  photon energy becomes comparable to the Rasha splitting near the
bottom of the conduction band. As a result, $\Delta\rho$
exhibits a strong dispersion caused by the coupling in \eq{curv-w}
between the two conduction subbands.

\subsection{Absolute sign of the optical rotation}
\label{sec:signs}

All gyrotropic effects have equal magnitudes and opposite signs for
two otherwise identical samples of opposite handedness.  Unfortunately
the experimental determination of the handedness is particularly
difficult for elemental crystals~\cite{tanaka-jpcm10}, and there are
conflicting claims in the literature as to which enantiomorph of
trigonal Te rotates the plane of polarization of light in which sense.

We are aware of three studies that tried to establish the handedness
of a Te sample, correlating it with the sign of the rotatory power
$\rho_0$.  The first work used etching techniques~\cite{koma-pss70},
the second polarized neutron diffraction~\cite{brown-ac96}, and the
third resonant x-ray diffraction~\cite{tanaka-jpcm10}.  In
Refs.~\cite{koma-pss70,brown-ac96} it was concluded that the plane of
polarization of light
rotates in the same sense as the bonded atoms in the spiral chains
(with our sign convention, that means $\rho_0>0$ for the right-handed
structure), and this has become the ``accepted wisdom'' in the
literature~\cite{fukuda-pss75,ades-josa75,stolze-pssb77,shalygin-pss12}.
However, the authors of the most recent study~\cite{tanaka-jpcm10}
arrived at the {\it opposite} conclusion, see
Erratum~\cite{tanaka-jpcm12}: right-handed Te has a negative $\rho_0$,
in agreement with our calculations.

Let us conclude with a comment on the sign of $\Delta\rho$
calculated in Refs.~\cite{vorobev-jetp79,shalygin-pss12} using a
$\mathbf{k\cdot p}$ model for the band-edge states. It was found in
those works that $\Delta\rho<0$ when the states with positive
(negative) $q_z$ at the top of the uppermost valence band are
dominated by atomic states with total angular momentum $j_z=-3/2$
($j_z=+3/2$). In Ref.~\cite{shalygin-pss12} that situation was assumed
to correspond to right-handed Te, since it leads to $\rho_0>0$ (see
Eq.~(15) in Ref.~\cite{ivchenko-spss75}), in agreement with
Refs.~\cite{koma-pss70,brown-ac96}.  However, an examination of our
{\it ab initio} results leads to the opposite conclusion. For example,
the lower panel of \fig{11}(c) shows that in right-handed Te the spin
magnetic moment of states near the top of the upper valence band is
negative for $q_z>0$. In an atomic picture, this corresponds to states
with total angular $j_z=+3/2$ being dominant at positive~$q_z$.
In conclusion, once the $\mathbf{k\cdot p}$ model is matched to our
{\it ab initio} wavefunctions it yields $\rho_0<0$ and $\Delta\rho>0$
for right-handed Te, in agreement with our calculations.\footnote{ The
  $\mathbf{k\cdot p}$ model of
  Refs.~\cite{vorobev-jetp79,shalygin-pss12} includes spin-orbit
  coupling in the valence bands only. This is an acceptable
  approximation, given that the spin-orbit induced Weyl points at the
  edge of the conduction band do not give a large contribution to the
  kFE at the CO$_2$ laser frequency. Recall from \sect{Te:PGE} that
  this was not the case for the CPGE: Without spin-orbit coupling in
  the conduction bands, the intrinsic part of the intraband CPGE would
  not change sign with temperature.}

\section{Kinetic magnetoelectric effect}
\label{sec:Te:kME}

\begin{figure}[t!]
\center
\includegraphics[width=\columnwidth]{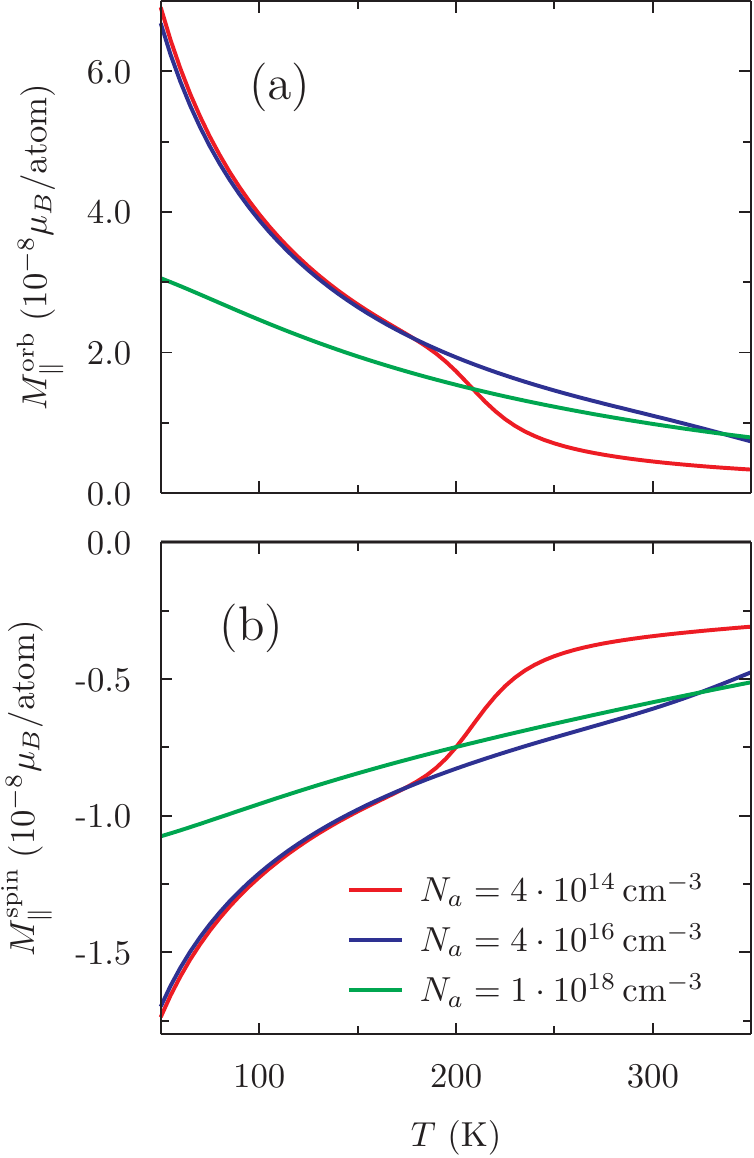}
\caption{Temperature dependence of the orbital and spin magnetization
  induced in right-handed Te, at different acceptor concentrations, by
  a current density $j_\|=1000$ A/cm$^2$.
  \label{fig:10}}
\end{figure}

\begin{figure*}[t!]
\center
\includegraphics[height=5.4cm]{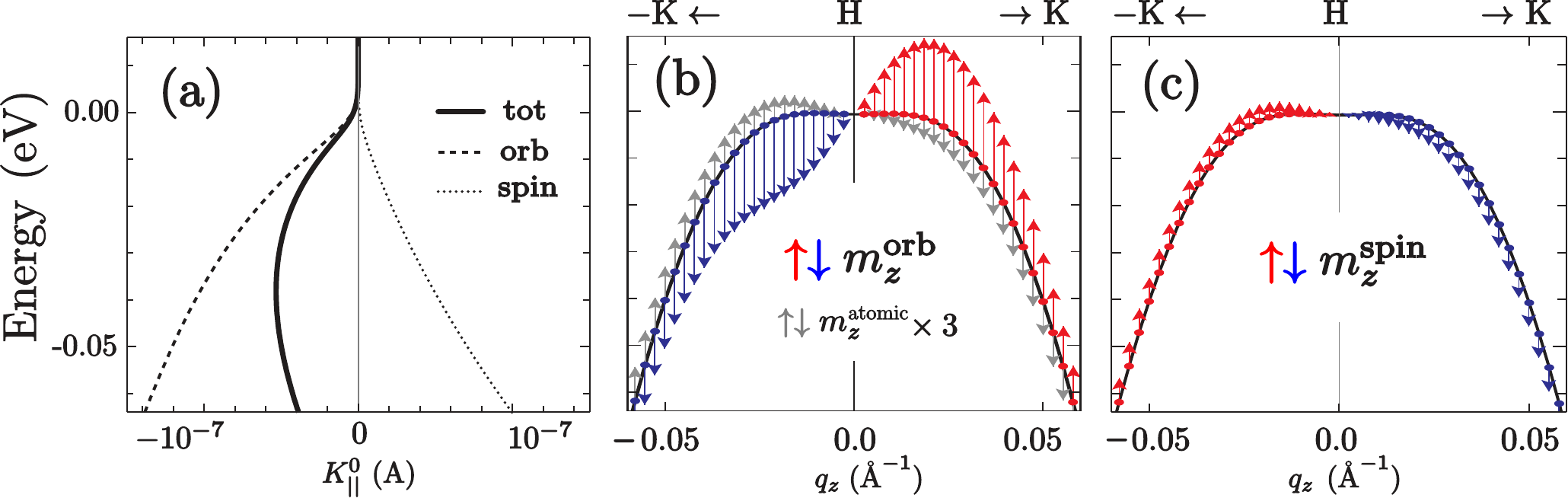}
\caption{ Microscopic mechanism of the kinetic magnetoelectric effect
  in right-handed Te. (a) Similar to the lower panel of \fig{8}(a),
  with $\wt D_\|^0$ replaced by $K_\|^{0}$ [the low-temperature limit
  of \eq{K_ab}].  The total $K_\|^0$ (heavy solid line) is decomposed
  into orbital (dashed line) and spin (dotted line) parts.  (b) and (c) Like
  the lower panels of \fig{8}(b), with $\wt\Omega_{\kk n}^z$ replaced
  by $m_{\kk n,z}^{\rm orb}$ [\eq{m-orb}] in (b), and by
  $m^{\rm spin}_{\kk n,z}=-\frac{1}{2}g_s\mu_{\rm B} \me{\psi_{\kk
      n}}{\sigma_z}{\psi_{\kk n}}$
  in (c) --- the orbital and spin parts of the intrinsic magnetic
  moment of a Bloch electron. In (b), the gray arrows denote orbital
  moments calculated according to \eq{m-atom}.  \label{fig:11}}
\end{figure*}

Along with the Faraday rotation of transmitted light, the flow of a
{\it dc} current through a gyrotropic crystal produces a macroscopic
magnetization.  So far, the intrinsic contribution to this effect has
only been calculated for model tight-binding
systems~\cite{yoda-sr15,yoda-arxiv17}. Our goal in this section is to
make quantitative estimates for $p$-Te, and to provide a microscopic
picture for the effect.

A current flowing along the trigonal axis induces a parallel
magnetization given by (Appendix~\ref{sec:kME})
\beq
\label{eq:kME-main}
M_\parallel=-\frac{K_\parallel j_\parallel}{2\pi C_\parallel}.
\eeq
The temperature and doping dependence of $M_\|$ calculated at
$j_\|=1000\text{ A/cm}^2$ is shown in \fig{10}.  In contrast to 2D
inversion layers where the current-induced magnetization is purely
spinlike~\cite{edelstein-ssc90,aronov-jetp89,ganichev-book12}, in
$p$-Te it has both orbital and spin components, shown separately in
\fig{10}. They have opposite signs and comparable magnitudes, with the
orbital effect being somewhat larger.\footnote{Without spin-orbit
  coupling the bulk kME would be purely
  orbital~\cite{yoda-sr15,zhong-prl16}, and we attribute the presence
  of a comparable spin contribution to the kME to the strong
  spin-orbit coupling in Te.}  Their magnitudes are
$\sim 10^{-8}\,\mu_\text{B}$/atom, six orders smaller than the
spontaneous orbital magnetization in bcc Fe~\cite{lopez-prb12} (recall
that comparable differences in orders of magnitude {relative to bcc
  Fe} were found earlier for the nonlinear AHE and for the kFE).

The current-induced spin density at 77~K,
$N_a=4\cdot 10^{16}\text{ cm}^{-3}$, and $j_\|=1400$ A/cm$^2$ was
estimated in Ref.~\cite{shalygin-pss12} to be $\sim$560
spins/$\mu$m$^3$ .  {Under the same conditions} our calculation yields
561 spins/$\mu$m$^3$, in a surprisingly perfect agreement. {While it
  may be difficult to directly measure such a small magnetization,
  indirect evidence for the kME in $p$-Te has already been gathered.
  In addition to the kFE~\cite{vorobev-jetp79,shalygin-pss12}, a
  current-induced splitting of nuclear magnetic resonance peaks was
  recently detected~\cite{furukawa-nc17}.}

{The dominance of the orbital contribution to $M_\|$ in \fig{10}
  implies that it remains positive over the entire temperature range.}
The signs of $M^{\rm orb}_\|$ and $M^{\rm spin}_\|$ can be understood
from \fig{11}. Panel~(a) shows the quantity $K^{0}_\|(\varepsilon)$
[defined {in terms of $K_\|$} in the manner of \eq{D-conv}] at the top
of the valence band, and the signs of its orbital and spin
contributions follow from panels (b) and (c), where it can be seen
that the $z$ component of the orbital (spin) moments of the band
states are antiparallel (parallel) to
$\partial E_{\kk n}/\partial k_z$.

The fact that the spin and orbital moments  are
antiparallel {for states} in the upper valence band is somewhat
surprising. Those states can be approximated as a linear combination
of atomic states with total angular momenta
$j_z=\pm 3/2$~\cite{ivchenko-jetp78,vorobev-jetp79,shalygin-pss12},
suggesting parallel spin and orbital moments.  This is confirmed by
the gray arrows in \fig{11}(b), which show the orbital moments
calculated {in the atomic-sphere approximation} as
\beq
\label{eq:m-atom}
m^{\rm atomic}_{\kk n,z}=-\mu_{\rm B}\sum_{ilm} m |\ip{u_{\kk n}}{ilm}|^2,
\eeq 
where the $\ket{ilm}$ are projectors onto spherical-harmonic states
localized on the $i$th atom in the unit cell, and
$\mu_{\rm B}=e\hbar/(2m_e)$ is the Bohr magneton.  As seen in
\fig{11}(b), the moments calculated from \eqs{m-orb}{m-atom} differ in
both sign and magnitude.  {This signals a breakdown of the atomic
  picture of orbital magnetism for states at the top of the valence
  band, highlighting the need to use the rigorous
  definition~(\ref{eq:m-orb}) of ${\bm m}_{\kk n}^{\rm orb}$ so as to
  include itinerant contributions related to the Berry curvature.}

In fact, the signs of $m_{\kk n,z}^{\rm orb}$ and
  $\Omega_{\kk n}^z$ are correlated for states in the upper valence
  band, as can be seen
by comparing the spectral decomposition of \eq{m-orb},
\beq
\label{eq:orb-spectral}
{\bm m}^\text{orb}_{\kk n}
=\frac{e}{2\hbar}\sum_m\left(E_{\kk m}-E_{\kk n}\right)
\im\left({\bm A}_{\kk nm}{\bm A}_{\kk mn}\right),
\eeq
with that of \eq{curv} [given by \eq{curv-w} at $\ww=0$], and
recalling from \sect{Faraday-T} that the upper valence band couples
most strongly to the lower conduction subbands, for which
$E_{\kk m}-E_{\kk n}>0$. {This analysis suggests that
  $m_{\kk n,z}^{\rm orb}$ and $\Omega_{\kk n}^z$ should be
  antiparallel, which is indeed the case: Compare \fig{11}(b) with
  \fig{3}(b).}


\section{Natural optical activity of doped tellurium\label{sec:Te:NOA-intra} }

The theoretical value of $\rho_0$ in Table~\ref{table:rotatory} was
calculated for undoped Te, and here we analyze how it changes under
doping. We consider two effects: the doping dependence of the
interband contribution, and the appearance in doped samples of an
intraband contribution, whose mechanism is closely related to that of
the kME~\cite{zhong-prl16,ma-prb15}. We calculate both effects at 77~K,
for the CO$_2$ laser frequency.

\begin{figure}[t!]
\includegraphics[width=\columnwidth]{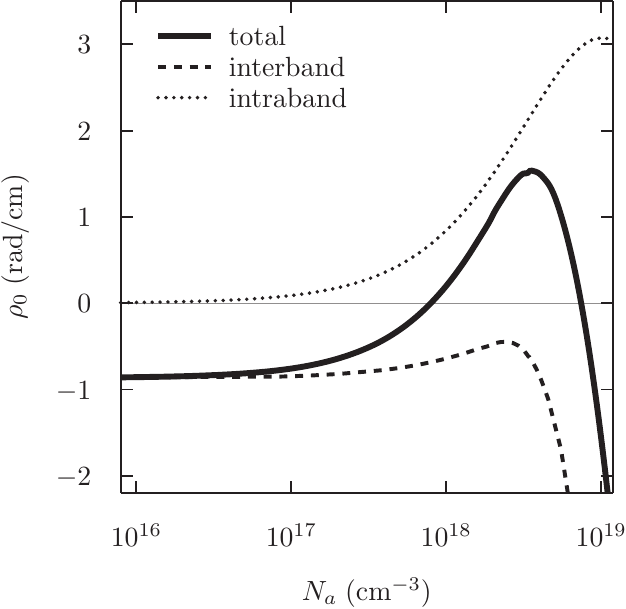}
\caption{Doping dependence of the natural rotatory power of
  right-handed Te at 77~K and $\hbar\ww=0.117$~eV, decomposed
  into interband and intraband contributions.\label{fig:12}}
\end{figure}

We begin with the doping dependence of the interband rotatory power, which can be taken into account by replacing
$\sum_{n,l}^{o,e}$ with $\sum_{n,l}f_{\kk n}(1-f_{\kk l})$ in \eq{gamma-inter}
(see Ref.~\cite{stolze-pssb77}).
As shown by the dashed line in \fig{12}, $\rho^{\rm inter}_0$
remain negative over the entire doping range.  At first its magnitude decreases
slightly with increasing $N_a$, due to a depopulation of the
upper valence band that blocks some of the interband
transitions~\cite{stolze-pssb77}. It reaches a minimum at
  $N_a\approx 2.5\cdot 10^{18}$~cm$^{-3}$, and then increases rapidly in magnitude. The rapid
increase is caused by transitions between the two upper valence bands,
which become possible at high doping~\cite{stolze-pssb77}. Although
the matrix elements for such transitions are
small~\cite{stolze-pssb77,shalygin-pss12}, along the HA line the band
separation is close to the CO$_2$ laser frequency of
$\hbar\ww=0.117\text{ eV}$ [see \fig{1}(b)], producing a resonant
enhancement.

We now turn to the intraband rotatory power, shown as the dotted
  line in \fig{12}. In Appendix~\ref{sec:NOA-intra} we obtained,
  following Refs.~\cite{zhong-prl16,ma-prb15},
\begin{subequations}
\label{eq:rho-intra}
\begin{align}
  \rho_0^{\rm intra}(\ww)&=
\frac{\ww^2\tau^2}{1+\ww^2\tau^2}\,\rho_0^{\rm clean},\\
\label{eq:rho-clean}
\rho_0^{\rm clean}&=
-\frac{4\pi\alfa}{ec}K_\perp,
\end{align}
\end{subequations}
with $K_\perp$ given by \eqs{K_ab}{K-Te}. 
Using the values of $\tau(N_a,T)$ from Ref.~\cite{peng-prb14}, we
conclude that up to $N_a=10^{20}$~cm$^{-3}$ the ``clean-limit''
condition $\ww\tau\gg 1$ is satisfied at the CO$_2$ laser frequency
and room temperature {(and below)}. Thus,
$\rho_0^{\rm intra}\approx\rho_0^{\rm clean}$ over the entire range of
\fig{12}. {$\rho_0^{\rm intra}$ has the opposite sign compared to
  $\rho_0^{\rm inter}$, and a negligible magnitude at} low doping.
 But while $|\rho_0^{\rm inter}|$
{initially} decreases as the doping level increases,
$|\rho_0^{\rm intra}|$ increases (more or less linearly) with $N_a$.
Interestingly, between $8\cdot 10^{17}$ and
$7.5\cdot 10^{18}$~cm$^{-3}$ the competition between {the two
  contributions}  results in a
sign reversal of $\rho_0$.

\section{Summary}

In summary, we have carried out a combined {\it ab initio} study of
several gyrotropic effects in $p$-doped Te.\footnote{{The
    computer code developed for this project was written as a module
    of the {\tt wannier90} package~\cite{Mostofi2008,Mostofi2014}, and will be made
    publicly available in a forthcoming release.}} The motivation was
provided by recent theoretical developments that recognized the
central role played by the Berry curvature and by the intrinsic
orbital moment in the description of such effects in the semiclassical
regime of low frequencies compared to the band splittings.  This
prompted us to revisit the pioneering infrared measurements of the
CPGE~\cite{asnin-jetp78,asnin-ssc79} and
kFE~\cite{vorobev-jetp79,shalygin-pss12} in bulk Te.

We found that the intrinsic mechanism for the intraband
CPGE~\cite{deyo-arxiv09,moore-prl10,sodemann-prl15} accounts for the
observed sign reversal of the CPGE with temperature, {and that} the
sign reversal is caused by the presence of Berry-curvature monopoles
(Weyl points) at the bottom of the conduction band.  This provides {an
  interesting} example of {the way in which} Weyl points can influence
physical observables in semiconductors.

Regarding the natural and current-induced optical rotation (kFE), our
calculations give rotatory powers whose magnitudes are within a factor
of two of the measured ones.  In agreement with
experiment~\cite{shalygin-pss12}, we find that $\Delta\rho$ and
$\rho_0$ have opposite signs when light propagates in the same
direction as the current.

As for the absolute sign of $\rho_0$, we find that in undoped samples
the plane of polarization rotates in the opposite sense to the bonded
atoms in the spiral chains.  This contradicts the result of early
attempts to determine the handedness of a Te
sample~\cite{koma-pss70,brown-ac96}, but agrees with the most recent
experimental determination~\cite{tanaka-jpcm10,tanaka-jpcm12}. We also
predict a sign reversal of $\rho_0$ over a significant doping range,
due to the competition between interband and intraband contributions
{to the natural optical activity}.

In order to compare our fully quantum-mechanical calculation of
$\Delta\rho$ with the semiclassical limit, the result was expressed in
terms of a quantity $\wt{\bm\Omega}_{\kk n}(\ww)$ that reduces to the
Berry curvature at $\ww=0$. We found that at the CO$_2$ laser
frequency, $\wt{\bm\Omega}_{\kk n}(\ww)$ at the top of the valence
band is very close to ${\bm\Omega}_{\kk n}$. Hence, the
low-temperature kFE is well described by the same Berry-curvature
parameter $D_\|$ that governs the intrinsic CPGE. This leads to a
definite sign relation between the two effects, which could be tested
by measuring both on the same sample.

We have also provided estimates for the magnitudes of other gyrotropic
effects that have not yet been observed, such as the nonlinear AHE and
the kME. Our estimates indicate that those effects are rather small in
$p$-Te. However, a recent study predicted a sizable nonlinear AHE in
Weyl semimetals~\cite{zhang-arxiv17}.

In closing, we hope that the present work will stimulate further
experimental and theoretical work exploring the role of the $k$ space
Berry curvature, intrinsic orbital moment, and Weyl points in
connection with gyrotropic effects in bulk crystals.

{{\it Note added:} After this work was submitted, a complementary
  theoretical study of the kME in $p$-Te
  appeared~\cite{pesin-arxiv2017}.  The authors used a $\bf k\cdot p$
  model to investigate extrinsic as well as intrinsic contributions to
  the kFE.  For doping concentrations up to a few $10^17 cm^{-3}$, 
  they find that the latter are dominant, with the same
  magnitude and sign as reported here.}

\begin{acknowledgments}

  We acknowledge support from Grant No.~FIS2016-77188-P from the
  Spanish Ministerio de Econom\'ia y Competitividad, Grant
  No. CIG-303602 from the European Commission, and from Elkartek Grant
  No. KK-2016/00025.  We would like to thank V.~A.~Shalygin for useful
  discussions.

\end{acknowledgments}

\section*{Appendices}
\appendix

In Appendices~\ref{sec:PGE} to \ref{sec:kME} we review the theory of
the various gyrotropic effects considered in the main text.  The
photogalvanic effect is treated in Appendix~\ref{sec:PGE}, the
nonlinear AHE in Appendix~\ref{sec:AHE}, optical rotation in
Appendix~\ref{sec:optical-rotation-all}, and the kinetic
magnetoelectric effect in Appendix~\ref{sec:kME}.
Concerning the microscopic theory of these effects, our aim is to
present a coherent picture based on a small number of basic
ingredients (Berry connections, curvatures, and intrinsic magnetic
moments).  We only consider the ``intrinsic'' contributions that can
be calculated from the electronic structure of the pristine crystal
supplemented by a phenomenological relaxation time $\tau$.  We
therefore neglect extrinsic effects due to skew-scattering and
side-jump processes at impurities~\cite{deyo-arxiv09,rou-prb17}.
Finally, Appendix \ref{sec:wannier} describes some technical details
of our Wannier-based numerical scheme.

\section{Photogalvanic effect}
\label{sec:PGE}

\subsection{Phenomenology}

Consider an oscillating electric field
\beq
\bE(\rr,t)=\re\left[\bE(\ww)e^{i(\qq\cdot\rr-\ww t)}\right].
\eeq
The current density induced at second order in the field amplitude
can be written
as~\cite{sodemann-prl15}
\begin{subequations}
\label{eq:pge+j-shg}
\begin{align}
\label{eq:j-tot}
j_a(t)&=\re\left(j_a^0+j_a^{2\ww}e^{-i2\ww t}\right),\\
\label{eq:pge}
j^0_a&=\frac{1}{2}\sigma_{abc}(\ww)\E_b(\ww)\E^*_c(\ww),\\
\label{eq:j-shg}
j^{2\ww}_a&=\frac{1}{2}\sigma_{abc}(\ww)\E_b(\ww)\E_c(\ww).
\end{align}
\end{subequations}
Equations~(\ref{eq:pge}) and~(\ref{eq:j-shg}) describe a {\it dc}
photocurrent and a second-harmonic current, respectively.

Writing $\sigma_{abc}=\lambda_{abc}+\gamma_{abc}$ and
$\E_b\E^*_c=\re\left(\E_b\E^*_c\right)+i\im\left(\E_b\E^*_c\right)$,
where the first and second terms in these expressions are, respectively,
symmetric and antisymmetric under $b\leftrightarrow c$, \eq{pge}
becomes
\beq
\label{eq:lpge-cpge}
j^0_a
=\frac{1}{2}
\left\{
  \lambda_{abc}\re\left(\E_b\E^*_c\right)
  -\gamma_{ab}\im\left({\bE}\times\bE^*\right)_b
\right\},
\eeq
where
$\gamma_{ab}=-i\eps_{bcd}\gamma_{acd}/2 =-i\eps_{bcd}\sigma_{acd}$/2.
The first (second) term describes the linear (circular) photogalvanic
effects. $\lambda_{abc}$ transforms 
like the piezoelectric tensor, and $\gamma_{ab}$ like the gyration
tensor~\cite{sturman-book92,ivchenko-book97}.

\subsection{Berry-curvature (``intrinsic'') contributions}

 The
intrinsic intraband contribution to the nonlinear conductivity
of a nonmagnetic crystal can be expressed in terms of the tensor $D$
in \eq{D_ab} as~\cite{sodemann-prl15}
\beq
\label{eq:sigma_abc}
\sigma_{abc}=-\frac{e^3\tau_\ww}{\hbar^2}\eps_{adc}D_{bd},
\eeq
where
\beq
\label{eq:tau-w}
\tau_\ww=\frac{\tau}{1-i\ww\tau}.
\eeq 
Combining Eqs.~(\ref{eq:pge+j-shg})--(\ref{eq:sigma_abc}) one finds
\begin{subequations}
\label{eq:j-lpge-cpge}
\begin{align}
\re\left(j^0_a\right)&=j_a^{\rm LPGE}+j_a^{\rm CPGE},\\
\label{eq:j-lpge}j^{\rm LPGE}_a&=
-\frac{e^3}{2\hbar^2}\re(\tau_\ww)
\eps_{adc}D_{bd}\re\left(\E_b\E_c^*\right),\\
\label{eq:j-cpge}
j_a^{\rm CPGE}&=
-\frac{e^3}{4\hbar^2}\im(\tau_\ww)
D_{ab}
\im
\left(\bE\times\bE^*\right)_b,
\end{align}
\end{subequations}
where LPGE stands for ``linear photogalvanic effect,'' and
$\text{Tr}(D)=0$ was used to eliminate one term from
\eq{j-cpge}~\cite{deyo-arxiv09}.

Consider the CPGE in trigonal Te with light propagating along the
trigonal axis.  Writing ${\bm q} =q_\parallel\hat{\bm z}$,
$\bE=\vert\bE\vert\ee$, and
$-\im\left({\bm e}\times{\bm e}^*\right)={\cal P}_{\rm circ}\hat{\bm
  q}$
where ${\cal P}_{\rm circ}$ is the degree of circular polarization,
and defining the intensity of incident light as
${I_0}=c\eps_ 0\vert\bE\vert^2/2$, \eq{j-cpge} becomes
\beq
\label{eq:cpge-Te}
j^{\rm CPGE}_{\parallel}=
\text{\rm sgn}(q_\parallel)
\left( 2\pi\alfa{\cal P}_{\rm circ}D_{\parallel} \right)
\im(\tau_\ww)\frac{e{I_0}}{\hbar},
\eeq
where $\alfa=e^2/(4\pi\eps_0\hbar c)$ is the fine-structure constant.
For positive helicity (${\cal P}_{\rm circ}>0$), the current flows
parallel (antiparallel) to the light beam when $D_{\parallel}>0$
($D_{\parallel}<0$).

\section{Nonlinear anomalous Hall effect}
\label{sec:AHE}

In the $\ww\rightarrow 0$ limit 
the total current from \eqs{pge+j-shg}{j-lpge-cpge} becomes
\beq
\label{eq:j-AH}
j_a=-\frac{e^3\tau}{\hbar^2}\eps_{adc}D_{bd}\E_b\E_c,
\eeq
with equal parts coming from the second-harmonic and LPGE currents
(the CPGE vanishes at $\ww\rightarrow 0$).  Since
$\bE\cdot\jj=0$, \eq{j-AH} describes a nonlinear anomalous
Hall current~\cite{sodemann-prl15}.

It is instructive to obtain \eq{j-AH} by replacing $f_0$ in
\eq{ahc+curv} with the change in the distribution function at linear
order in an applied static field, 
\beq
\label{eq:Delta-f}
\Delta f\equiv f-f_0
=-e\tau\bE\cdot{\bm v}_{\kk n}
\left(-\frac{\partial f_0}{\partial E}\right)_{E=E_{\kk n}}.
\eeq
Doing so yields
\beq
\label{eq:delta-ahc}
\Delta\sigma^{\rm A}_{ab}=\frac{e^3\tau}{\hbar^2}\eps_{abd}D_{cd}\E_c
\eeq
for the field-induced AHC, in agreement with \eq{j-AH}.

Inserting \eq{D-Te} for the tensor $D$ in Te into \eq{j-AH} for the current, we obtain
\begin{subequations}
\begin{align}
j_x&=\frac{3e^3\tau}{2\hbar^2}D_\parallel\E_z\E_y,\\
j_y&=-\frac{3e^3\tau}{2\hbar^2}D_\parallel\E_z\E_x,\\
j_z&=0.
\end{align}
\end{subequations}
The nonlinear current flows in the plane perpendicular to the trigonal
axis, and the effect can be viewed as an in-plane linear AHE induced
by the out-of-plane field component $\E_\parallel\equiv \E_z$. The
effective field-induced AHC is
\beq
\label{eq:sigma-AH-eff-a}
\sigma^{\rm A}_{xy}(\E_\parallel)
=\frac{3e^3\tau}{2\hbar^2}D_\parallel \E_\parallel
=\frac{3e^3}{2\hbar^2}\frac{D_\parallel}{\sigma_\parallel/\tau} j_\parallel,
\eeq
where in the second equality we inverted Ohm's law to express the
result in terms of $j_\parallel$ (the nonzero components of the Ohmic
conductivity are $\sigma_\perp\equiv\sigma_{xx}=\sigma_{yy}$ and
$\sigma_\parallel\equiv\sigma_{zz}$).  In the constant relaxation-time
approximation we have $\sigma_\|/\tau=(2\pi e/\hbar)C_\|$, with
\beq
\label{eq:C_ab}
C_\|=\frac{e}{h}\int\dk\sum_n\,
\left(\frac{\partial E_{\kk n}}{\partial{k_z}}\right)^2
\left(-\frac{\partial f_0}{\partial E}\right)_{E=E_{\kk n}}
\eeq
a positive quantity with
units of surface current density. With this notation,
the current-induced AHC reads
\beq
\label{eq:sigma-AH-eff}
\sigma^{\rm A}_{xy}(j_\parallel)
= 
(e^2/h)(3D_\parallel/2) (j_\parallel/C_\parallel),
\eeq
where $e^2/h$ is the quantum of conductance, $D_\parallel$ is
dimensionless, and $j_\parallel/C_\parallel$ has units of inverse
length.

\section{Optical rotation}
\label{sec:optical-rotation-all}

\subsection{Phenomenology}
\label{sec:optical-rotation-phen}

The dielectric tensor of trigonal Te has the
form~\cite{shalygin-pss12}
\beq
\label{eq:eps-j-Te} {\bm\vareps}(\ww,q_\|,j_\parallel)= \left(
\begin{array}{ccc}
\vareps_\perp & \vareps^{\rm A}_{xy}(\ww,q_\|,j_\parallel) & 0\\
-\vareps^{\rm A}_{xy}(\ww,q_\|,j_\parallel) & \vareps_\perp & 0\\
0 & 0 & \vareps_\parallel
\end{array}
\right).
\eeq
In equilibrium, the antisymmetric part $\vareps^{\rm A}_{xy}$
responsible for optical rotation is linear in the wavevector $q_\|$ of
light propagating inside the crystal along the trigonal axis.
Under a steady current flow, $\vareps^{\rm A}_{xy}$ acquires a new
contribution closely related to the nonlinear AHC of
Appendix~\ref{sec:AHE}. It is linear in $j_\parallel$ and zeroth-order
in $q_\|$, giving rise to the kFE.  (As for the diagonal elements
$\vareps_\perp$ and $\vareps_\parallel$, they are independent of
$j_\parallel$ and $q_\|$ to linear order.)

Before proceeding further, let us specify our sign convention for
optical rotation.
We say that the rotatory power $\rho$
is positive when the sense of rotation of the electric field
vector is counterclockwise as seen by an observer looking toward the
light source.
With this choice we have\footnote{Compare with Eq.~(2) in Ch.~XIV of
  Ref.~\cite{nye-book85}, where the {\it opposite} sign convention for
  $\rho$ was adopted. Therein, ``left-circular polarization'' refers
  to our positive helicity (see also Ref.~\cite{jackson-book99}).}
\beq
\label{eq:rho-def}
\rho
=\frac{\pi}{\lambda_0}\re\left(n_--n_+\right)
=\frac{\ww}{2c}\re\left(n_--n_+\right),
\eeq
where $\lambda_0$ is the wavelength in vacuum, and $n_+$ and $n_-$ are
the complex indices of refraction for circularly polarized waves of
positive and negative helicity, respectively, with polarization vectors given by
\beq
{\bm e}_\pm=
\frac{\hat{\bm x}\pm i{\rm sgn}(q_\parallel)\hat{\bm y}}{\sqrt{2}}.
\eeq

Assuming a sufficiently small current density such that
$\vert\vareps^A_{xy}/\vareps_\perp\vert\ll 1$, one
finds~\cite{liu-book16,shalygin-pss12}

\beq
\label{eq:Delta-n}
n_--n_+\approx -{\rm sgn}(q_\parallel)
i\frac{\vareps^{\rm A}_{xy}/\vareps_0}{n_\perp},
\eeq
where $n_\perp\equiv \sqrt{\vareps_\perp/\vareps_0}$.
Converting to conductivities using
\beq
\label{eq:epsilon-sigma}
\vareps_{ab}(\ww)=\vareps_0
\left[
  \delta_{ab}+\frac{i}{\ww\vareps_0}\sigma_{ab}(\ww)
\right],
\eeq
we obtain 
\beq
\label{eq:rho}
\rho(\ww,j_\parallel)={\rm sgn}(q_\parallel)
\frac{\re\,\sigma^{\rm A}_{xy}(\ww,j_\parallel)}
     {2c\vareps_0n_\perp(\ww)}
\eeq
at nonabsorbing frequencies, with
\beq
\label{eq:n-perp}
n_\perp(\ww)
=\left[
  1-\frac{1}{\ww\vareps_0}\im\,\sigma_\perp(\ww)
\right]^{1/2}.
\eeq

In the following, we expand the rotatory power
as~\cite{shalygin-pss12}
\beq
\label{eq:rho-expansion}
\rho(\ww,j_\parallel)=\rho_0(\ww)+
\Delta\rho(\ww,j_\|)+{\cal O}\left(j_\|^2\right).
\eeq
$\rho_0$ is the natural rotatory power at $j_\|=0$, and
$\Delta\rho(j_\|)$ is the change in rotatory power at linear
order in $j_\|$.

\subsection{Natural optical rotation}
\label{sec:NOA}

Natural optical rotation is described by
$\sigma^{\rm A}_{xy}(\qq,\ww)$ at first order in $q_z$, which is
conventionally written as~\cite{landau-EM}
\beq
\sigma^{\rm A}_{xy}(\ww,\qq)=\ww\varepsilon_0\gamma_{xyz}q_z
=\text{sgn}(q_\parallel)\ww\varepsilon_0\gamma_{xyz}
\vert q_\parallel\vert,
\label{eq:epsilon-gamma}
\eeq
where $\gamma_{xyz}$ has units of length.  Using
$\vert q_\parallel\vert/\re\,n_\perp=\ww/c$, \eq{rho}
becomes~\cite{ivchenko-spss75}
\beq
\label{eq:rho-c}
\rho_0(\ww)=\frac{\ww^2}{2c^2}\re\,\gamma_{xyz}(\ww).
\eeq
 Note that
the natural rotatory power does not reverse sign with $q_\parallel$.
Thus, if a linearly-polarized ray travels back and forth inside the
material the plane of polarization is unchanged when it returns to the
initial point~\cite{landau-EM,liu-book16}.

We now turn to the microscopic theory.  The natural optical activity
of nonconducting crystals is governed by virtual interband
transitions~\cite{natori-jpsj75,ivchenko-spss75,malashevich-prb10},
and the rotatory power decreases as $\ww^2$ at frequencies well
  below those of interband transitions.  Instead, conducting crystals remain
optically active at such low frequencies due to intraband
processes~\cite{zhong-prl16,ma-prb15}. {Thus, the rotatory power
  of a conducting crystal is given by}
\beq \rho_0{(\ww)}=\rho_0^{\rm inter}{(\ww)}+\rho_0^{\rm
  intra}{(\ww)}.  
\eeq
In the following, both contributions are calculated.

\subsubsection{Interband natural optical rotation\label{sec:NOA-inter}}

Following Ref.~\cite{malashevich-prb10} we write,
with $\partial_c\equiv\partial/\partial k_c$,
\begin{multline}
\re\,\gamma_{abc}^{\mathrm{inter}}(\ww)=\frac{e^2}{\varepsilon_0\hbar^2}
\int[d\kk]
\sum_{n,l}^{o,e}\,
\Bigl[ \\
\frac{1}{\ww_{ln}^2-\ww^2} 
\re\left(A_{ln}^bB_{nl}^{ac}-A_{ln}^aB_{nl}^{bc}\right) \\
-\frac{3\ww_{ln}^2-\ww^2}{(\ww_{ln}^2-\ww^2)^2} 
\partial_c(E_l+E_n)\im\left(A_{nl}^aA_{ln}^b\right)   
\Bigr].
\label{eq:gamma-inter}
\end{multline}
The summations over $n$ and $l$ span the occupied ($o$) and empty
($e$) states respectively, $\ww_{ln}=(E_l-E_n)/\hbar$, and we omit the ${}_\kk$ subscript for brevity.
Here
\beq
\label{eq:A}
A^a_{ln}=i\ip{u_l}{\partial_a u_n}
\eeq
is the matrix generalization of the Berry connection appearing
  in \eq{curv}.  Finally, the matrix
$B_{nl}^{ac}$ has both orbital and spin contributions given by
\beq
\label{eq:B-ac-orb}
B_{nl}^{ac\,({\rm orb})}=
  \bra{u_n}(\partial_aH)\ket{\partial_c u_l}
 -\bra{\partial_c u_n}(\partial_aH)\ket{u_l}
\eeq
and
\beq
\label{eq:B-ac-spin}
B_{nl}^{ac\,({\rm spin})}=-\frac{i\hbar^2}{m_e}\epsilon_{abc}
\bra{u_n}\sigma_b\ket{u_l}.
\eeq
In Te the spin matrix elements contribute less than 0.5\% of the total
$\rho_0^{\rm inter}$, {and can be safely ignored}.  Writing
$H=\sum_m \ket{u_m} E_m \bra{u_m}$, the orbital matrix elements become
\bea B_{nl}^{ac\,({\rm
    orb})}&=&-i\partial_a(E_n+E_l)A_{nl}^c\nn &+&\sum_m \Bigl\{
(E_n-E_m) A_{nm}^aA_{ml}^c \nn &&\,\,\,\,\,\,\,\,\,-(E_l-E_m)
A_{nm}^cA_{ml}^a \Bigr\}.
\label{eq:Bnl-sum}
\eea
This reduces the calculation of $B^{\text{(orb)}}$ to the evaluation
of band gradients and off-diagonal elements of the Berry connection
matrix, and both operations can be carried out efficiently in a
Wannier-function basis~\cite{yates-prb07}.

In our implementation, the summation in \eq{Bnl-sum} is restricted
to 
the $s$ and $p$ bands included in the Wannierization procedure (see
Appendix \ref{sec:wannier}). To check how quickly the calculated
$\rho_0^{\rm inter}$ converges with the number of bands, we redid the
calculation keeping only the four bands (two valence and two
conduction) closest to the gap, and found that
the value changed by only 10\% compared to a calculation including all
$s$ and $p$ states.  This is consistent with the conclusion of
Ref.~\cite{stolze-pssb77} that the natural optical activity of Te is
contributed mainly by transitions between states near the energy gap.

  \subsubsection{Intraband natural optical
    rotation \label{sec:NOA-intra}}

  Here we calculate $\rho_0^{\rm intra}$ following
  Refs.~\cite{zhong-prl16,ma-prb15}.  Combining Eqs.~(5a) and (S61) in
  Ref.~\cite{zhong-prl16} and noting that in our notation the tensor
  $\alpha^{\rm GME}$ defined therein is given by
  $-i\ww(e/\hbar)\tau_\ww K$, we find
\beq
\re\,\gamma_{abc}^{\mathrm{intra}}(\ww)=\frac{e\im\,\tau_\ww}{\ww\vareps_0\hbar}
\left(\eps_{acd}K_{bd}-\eps_{bcd}K_{ad}\right).
\eeq
Using \eq{K-Te} for the tensor $K$ in Te leads to
\beq
\re\,\gamma_{xyz}^{\mathrm{intra}}(\ww)=-\frac{2e\im\,\tau_\ww}{\ww\vareps_0\hbar}K_\perp.
\eeq
The intraband rotatory power of \eq{rho-intra} is obtained by
inserting this expression in \eq{rho-c}.

\subsection{Current-induced optical rotation \label{sec:eps-j}}

Let us now obtain a microscopic expression for $\Delta\rho$ in
\eq{rho-expansion}, by expanding \eq{rho} to first order in
$j_\|$. For that purpose, it is sufficient to expand the tensor
$\re\,\sigma^{\rm A}_{xy}(\ww)$ in the numerator. At $j_\|=0$ {it
  is given by the following finite-frequency generalization of
  \eq{ahc+curv},}
%
\beq
\label{eq:sigma-w-0}
\re\,\sigma^{\rm A}_{ab}(\ww)=
-\frac{e^2}{\hbar}\int\dk\sum_n
f_0(E_{\kk n})\eps_{abc}\wt\Omega^c_{\kk n}(\ww),\\
\eeq
{where the quantity}
%
\beq
\label{eq:curv-w}
\wt{\bm\Omega}_{\kk n}(\ww)=-
\sum_m\,\frac{\ww^2_{\kk mn}}{\ww^2_{\kk mn}-\ww^2}
\im\left({\bm A}_{\kk nm}\times{\bm A}_{\kk mn}\right)
\eeq
{reduces to the Berry curvature at $\ww=0$.}\footnote{Contrary to
  the Berry curvature, the divergence of
  $\tilde{\bm\Omega}_{\kk n}(\ww)$ is generally nonzero. As a result,
  $\tilde D(\ww)$ given by \eq{D-tilde} can have a nonzero trace at
  finite frequencies, i.e., $\tilde{D}_\|\neq-2\tilde{D}_\perp$ in
  Te.}

The correction to \eq{sigma-w-0} at first order in $j_\|$ can be
obtained by replacing $f_0$ therein with $\Delta f$ given by
\eq{Delta-f}. Following Appendix~\ref{sec:AHE} we obtain 
\beq
\re\,\sigma^{\rm A}_{xy}(\ww,j_\parallel)
=(e^2/h)\wt{D}_\parallel(\ww)(j_\parallel/C_\parallel) 
\eeq
with $\wt{D}_\parallel(\ww)$ given by \eq{D-tilde}, and inserting this
expression in \eq{rho} we arrive at \eq{rho-j-main} for $\Delta\rho$.
Note that $\Delta\rho$ reverses sign with $q_\parallel$, contrary to
$\rho_0$: Like the conventional Faraday
effect~\cite{landau-EM,liu-book16}, the kFE is nonreciprocal.

The final step is to determine the refraction index $n_\perp$
appearing in \eq{rho-j-main}. For that purpose, we evaluate the
quantity $\im\,\sigma_\perp(\ww)$ in \eq{n-perp} using
\bea
\label{eq:sigma-S}
\im\,\sigma_\perp(\ww)&=&-\frac{e^2}{\hbar}\int\dk
\sideset{}{'}\sum_{nm}f_0(E_{\kk n})
\left[1-f_0(E_{\kk m})\right]\nn
&\times&\frac{\ww^2_{\kk mn}}{\ww^2_{\kk mn}-\ww^2}
\left( \vert A_{\kk nm}^x\vert^2+ \vert A_{\kk nm}^y\vert^2 \right),
\eea
where the prime on the summation indicates that the term $m=n$ is
excluded.  This expression gives the interband contribution to
$\im\,\sigma_\perp(\ww)$. Since at the CO$_2$ laser frequency we have
$\ww\tau\gg 1$ across the entire range of temperatures and doping
levels considered in our calculations, the intraband (Drude)
contribution is negligible.

\section{Kinetic magnetoelectric effect}
\label{sec:kME}

The kME effect in a conducting gyrotropic crystal is described
phenomenologically by~\cite{zhong-prl16}
\begin{subequations}
\label{eq:kME}
\begin{align}
j_a^\BB(\ww)&=i\ww\alpha_{ab}(\ww)B_b(\ww),\\
M_a(\ww)&=\alpha_{ba}(\ww)\E_b(\ww).
\end{align}
\end{subequations}
In the limit $\ww\tau\ll 1$ 
where $\alpha_{ab}(\ww)$ becomes real 
we have
\begin{subequations}
\label{eq:kME-lowfreq}
\begin{align}
j^\BB_a(t)&=-\alpha_{ab}(0)\dot{B}_b(t),\\
\label{eq:M-E-0}
M_a(t)&=\alpha_{ba}(0)\E_b(t),
\end{align}
\end{subequations}
which for an isotropic gyrotropic medium
($\alpha_{ab}=\alpha\delta_{ab}$) reduces to Eqs.~(1) and ~(3) of
Ref.~\cite{levitov-jetp85}.

It is convenient to introduce a reduced (dimensionless)
magnetoelectric tensor
\beq
\label{eq:alpha}
\alpha^{\rm r}_{ab}(\ww)=c\mu_0\alpha_{ab}(\ww),
\eeq
in direct analogy with the standard description of magnetoelectric
couplings in insulators~\cite{coh-prb11}.  The intrinsic part is given
in terms of \eqs{D_ab}{tau-w} by
\beq
\label{eq:alpha-r}
\alpha^{\rm r}_{ab}(\ww)=-4\pi\alfa\frac{\tau_\ww}{e}K_{ab}.
\eeq
It can be verified that at $\ww=0$ this expression agrees with
that obtained in Ref.~\cite{yoda-sr15} for the magnetization induced
by a static $\bE$ field.  Specializing to Te and following
Appendix~\ref{sec:AHE} to recast the result in terms of $j_\|$, we
obtain
\beq
\label{eq:M-j-a}
M_\parallel
=(e/8\pi^2\alfa)(\alpha^{\rm r}_\parallel(0)/\tau)
(j_\parallel/C_\parallel),
\eeq
which combined with \eq{alpha-r} becomes \eq{kME-main}.

\section{\label{sec:wannier} Wannier interpolation}

In order to interpolate in $k$ space the energy bands and other
quantities (see below), we use the formalism of maximally-localized
Wannier functions~\cite{PhysRevB.56.12847,PhysRevB.65.035109}, as
implemented in the {\tt Wannier90} code
package~\cite{Mostofi2008,Mostofi2014}.  We construct four
disentangled Wannier functions per tellurium atom and per spin
channel, for a total of 24 Wannier functions per cell.  The $5s$ and
$5p$ bands of trigonal Te are well separated from the lower $d$
states, and they cross with higher-lying sates only in a small region
of the Brillouin zone. Thus we set the outer energy window for the
disentanglement procedure~\cite{PhysRevB.65.035109} from -20 to
+5~eV relative to the valence-band maximum, so as to cover all $s$ and
$p$ bands.  The inner frozen window spans the range from -20 to
+2.5~eV, and we choose atom-centered $sp^3$-type trial orbitals for
the initial projections.  This choice of Wannier functions differs
from that of Ref.~\cite{tsirkin-prb17,hirayama-prl15}, where only $5p$
states were included in the wannierization.

The Wannier basis is also used to evaluate the $k$ space
  quantities entering the expressions for the response tensors,
  namely: the band gradient ${\bm\nabla}_\kk E_{\kk n}$, the Berry
  curvature ${\bm\Omega}_{\kk n}$ [\eq{curv}], the intrinsic orbital
  moment [\eq{m-orb}], and the off-diagonal elements of the Berry
  connection matrix ${\bm A}_{\kk nm}$ [\eq{A}]. The Wannier
  interpolation of these quantities is described in
  Refs.~\cite{wang-prb06,yates-prb07,lopez-prb12}.

  When evaluating the response tensors, the integrations over the
  Brillouin zone are performed using a uniform grid of
  $200\times200\times200$ $k$ points.  In the case of responses that
  can be expressed in the form of \eq{D-conv}, when $\varepsilon$ is
  close to the band gap (no further than 100 meV from the band edges),
  only $k$ points in the vicinity of H and H' contribute, due to the
  factor $(-\partial f_0/\partial E)$ in that equation.  In such
  cases, we use a grid of $200\times200\times200$ $k$ points within a
  small box centered at~H that amounts to less than 0.2\% of the
  entire Brillouin zone, and then multiply the result by two in order
  to account for H'. This allows us to increase the numerical accuracy
  for $\varepsilon$ near the band gap, which is the energy range that
  contributes most the response.


%

\end{document}